\documentclass[aps,twocolumn,floatfix,footinbib,groupedaddress,superscriptaddress,showpacs]{revtex4-1}

\usepackage{amsmath,amsfonts}
\usepackage{graphicx}
\usepackage{epsfig}
\usepackage[breaklinks=true,colorlinks=true,linkcolor=blue,urlcolor=blue,citecolor=blue]{hyperref}

\usepackage[printonlyused]{acronym}
\usepackage{multirow}
\usepackage{color}
\usepackage{listings}
\usepackage{color}
\usepackage{listings}
\usepackage{bm}
\definecolor{mygreen}{rgb}{0,0.6,0}
\definecolor{mygray}{rgb}{0.5,0.5,0.5}
\definecolor{mymauve}{rgb}{0.58,0,0.82}
\usepackage[normalem]{ulem} 
\usepackage{soul} 

\renewcommand{\vec}[1]{\boldsymbol{#1}}

\newcommand{\matnew}[1]{\mathbf{#1}}

\newcommand{\transpose}{\mathrm{t}}
\newcommand{\diag}{\mathrm{diag}}

\newcommand{\ignore}[1]{}

\begin{document}

\title{Linear-algebraic bath transformation for simulating complex open quantum systems}


\author{Joonsuk Huh}
\email{Email: huh@fas.harvard.edu}
\affiliation{Department of Chemistry and Chemical Biology, Harvard University, Cambridge, Massachusetts 02138, United States}
\author{Sarah Mostame}
\affiliation{Department of Chemistry and Chemical Biology, Harvard University, Cambridge, Massachusetts 02138, United States}
\author{Takatoshi Fujita}
\affiliation{Department of Chemistry and Chemical Biology, Harvard University, Cambridge, Massachusetts 02138, United States}
\author{Man-Hong Yung}
\affiliation{Center for Quantum Information, Institute for Interdisciplinary Information Sciences,
Tsinghua University, Beijing, 10084, People's Republic of China}
\author{Al\'an Aspuru-Guzik}
\email{Email: aspuru@chemistry.harvard.edu}
\affiliation{Department of Chemistry and Chemical Biology, Harvard University, Cambridge, Massachusetts 02138, United States}

\date{\today\ }





\begin{abstract}
In studying open quantum systems, the environment is often approximated as a collection of non-interacting harmonic oscillators, a configuration also known as the star-bath model. 
It is also well known that the star-bath can be transformed into a nearest-neighbor interacting chain of oscillators.
The chain-bath model has been widely used in renormalization group approaches.
%
%
The transformation can be obtained by recursion relations or orthogonal polynomials.
%
Based on a simple linear algebraic approach, we propose a bath partition strategy to reduce the system-bath coupling strength. 
As a result, the non-interacting star-bath is transformed into a set of weakly-coupled multiple parallel chains. 
The transformed bath model allows complex problems to be practically implemented on quantum 
simulators, and it can also be employed in various numerical simulations of open quantum dynamics.
\end{abstract}

\maketitle



\section{Introduction}
\begin{figure*}[htb]
\begin{center}
\resizebox{\linewidth}{!}{\includegraphics{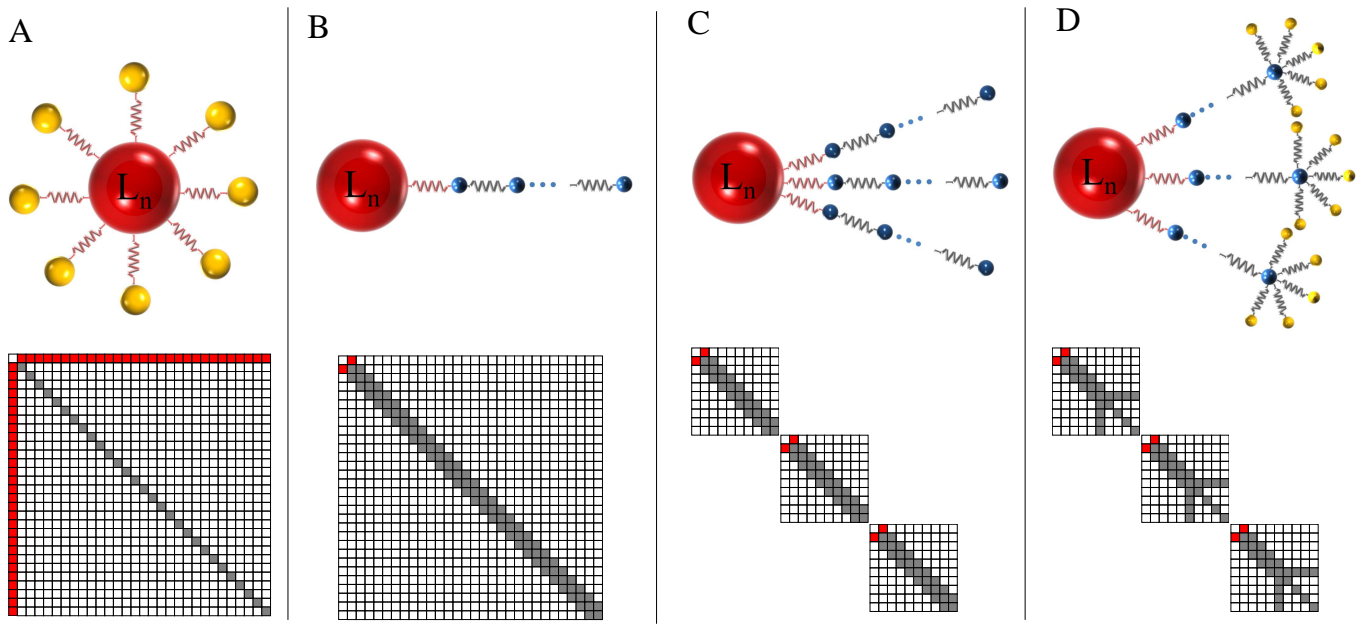}}
\caption{{\bf Top panels:} Various harmonic oscillator bath models. The red spheres represent system operators and the couplings are shown as springs. The couplings of primary modes, which are directly coupled to the system, are indicated in red. The yellow and blue spheres represent bath oscillators. 
{\bf A}.~{\em Star-bath model}: non-interacting quantum harmonic oscillators (yellow) are coupled to a system operator (red). 
{\bf B}.~{\em Chain-bath model}: a system operator (red) is coupled to a single interacting bath oscillator chain (blue). 
{\bf C}.~{\em Multiple-chain-bath model}: a system operator (red) is coupled to multiple interacting bath oscillator chains (blue). 
{\bf D}.~{\em Star-chain bath model}: a system operator (red) is coupled to multiple chains of bath operators (blue) and final bath oscillators are coupled to non-interacting bath oscillators (yellow).
{\bf Down panels:} The matrix representations (Eq.~\ref{eq:GSHU}) are shown for the corresponding top panel diagrams. The primary mode couplings to the system are given in red squares and other non-zero elements are shown in gray squares. The first column and row of each matrix correspond to the primary system-bath couplings. The diagonal elements, except the first one, are the frequencies of the bath oscillators. 
}\label{fig:bathmodel}
\end{center}
\end{figure*}

Problems associated with open quantum systems are of interest in various research fields~\cite{BRE02}.
%
%
In the theory of open quantum systems, the universe is partitioned into system and bath components. 
The system of interest is then coupled to the bath degrees of freedom (DOF) by means of an effective Hamiltonian. Solving the full quantum dynamics with currently known exact analytic or numerical methods are not feasible as the system and bath DOF increase. 
A simple but effective approach to model these vibrations is to treat them as a collection of non-interacting quantum harmonic oscillators bilinearly coupled to the system~\cite{BRE02,May:1320614}. 
The system-bath interaction is then characterized by a spectral density function (SDF) that represents the coupling strength in the frequency domain~\cite{BRE02,May:1320614}.
The energy transfer in photosynthetic systems is an example of a complex open quantum system, where the pigments involved in the energy transfer interact with a richly-structured set of molecular vibrations, and hence a very structured SDF~\cite{grondelle00}.

The spin-boson model~\cite{BRE02} is one of the simplest models for studying the dynamics of open quantum systems. 
In the most common representation, the spin-boson model is mathematically represented as a set of non-interacting oscillators coupled to the system. This can be graphically-represented in a star configuration as shown in Fig.~\ref{fig:bathmodel}{\bf A}~\cite{Bulla2003,Bulla2005}.
Generalized spin-boson models, such as the Hubbard-Holstein model, have been successfully employed to describe the energy transfer process in  photosynthetic antenna complexes~\cite{Huh:2014}. Some numerically-exact methods have been so far developed to solve these models, see, for example; 
reduced-density-matrix approaches, such as hierarchy equations of motion (HEOM)~\cite{Ishizaki2005,Ishizaki2009,Kreisbeck2011}, stochastic approaches~\cite{Strunz1999,Lacroix2008,Li2011,Orth2013}, multi-configuration time-dependent Hartree~\cite{Wang2008}, numerical renormalization group (NRG)~\cite{Bulla2003,Bulla2005,Schollw_ck_2005,Bulla_2008,Prior_2010,Chin_2011}, and path-integral approaches~\cite{Nalbach2011}, amongst many others. However, the applicability of numerically-exact methods is limited by the system size and the bath DOF. For example, the simulation of HEOM with current computers is limited to $\approx$40 sites of the system with only a single Drude-Lorentzian peak representing the bath~\cite{Kreisbeck2011,Mostame:2014}.
%
In the other hand, the renormalization group approach~\cite{Schollw_ck_2005,Chin_2010,Chin_2011} could be used for relatively large systems. 
However, the system and bath size that can be handled is still far from that required for solving problems at biological scales. 
%
In this approach, then the bath transformation from the non-interacting bath model (Fig.~\ref{fig:bathmodel}{\bf A}) to the 1-D Wilson chain (Fig.~\ref{fig:bathmodel}{\bf B})~\cite{Wilson_1975} is necessary.
The collective modes of the bath oscillators in the chain model has been used widely in various fields such as quantum molecular dynamics~\cite{Garg_1985,Tretiak_1996}, open quantum dynamics~\cite{Mori_1965,Dupuis_1967,Grigolini_1982,Burghardt2012,Chenel2014}, quantum information~\cite{Skinner2008} and nuclear physics~\cite{Iachello_1987,Caprio_2005}.

Quantum simulators defined as controlled quantum devices that can effectively reproduce the dynamics of any other quantum system~\cite{Feynman1982,Feynman1986,Lloyd1996,Buluta2009} could become an attractive alternative for solving the dynamics of open quantum systems "directly". Different platforms can be used for implementing quantum simulators, such as superconducting qubits~\cite{Skinner2008,Houck2012,Mostame2012,Ballester2012,Mei2013,Heras2014,Vladimir2014,Mostame:2014,Pedram2014},
trapped ions~\cite{Cirac2000,Ralf2007,Gerritsma2011,Casanova2011a,Casanova2011b,Casanova2012a,Blatt2012,Vladimir2012,Schindler2013}, quantum optics~\cite{Lanyon2010,Ralf2011,Herrera2011,Aspuru2012,Bloch2012}, nuclear magnetic resonance (NMR)~\cite{Du2010,Lu2011,Li2011,Zhang2011,Feng2013} or a system of electrons~\cite{Mostame2004,Mostame2008}.
%
%

The experimental implementation of quantum simulators of open quantum system dynamics, for example, using superconducting circuits~\cite{Mostame2012,Mostame:2014}, poses challenges due to at least two of the main current constraints in the realizable circuits.
First, the number of quantum bath oscillators, that are directly coupled to a system operator (qubit), is limited by the physical size of the superconducting loop that embodies the qubit. Hence, a star-model approach with many oscillators coupled to the qubit may pose fabrication challenges. A physical layout that involves less oscillators directly coupled to a qubit is more experimentally realizable~\cite{Mostame2012}. In addition, the coupling strength of the qubit to the bath may be limited. In superconducting qubits, the system-bath coupling strength should not exceed a certain percentage of the frequency of the quantum oscillator~\cite{Mostame:2014}. 

%
%
%
%
In this work, we address the question of how the two mentioned implementation issues can be resolved by a unitary bath transformation which introduces interaction terms among the transformed quantum oscillators. 
In chain-like bath models 
(Fig.~\ref{fig:bathmodel}{\bf B})
~\cite{Chernyak1996,Chou2008,Skinner2008,Bulla2003,Bulla2005,Chin_2010,Prior_2010,Burghardt2012,Chenel2014}, only one bath oscillator is directly coupled to the system. However, in some cases, one needs to couple more than a single chain to deal with the limitation of the oscillator-qubit coupling mentioned above.
Here, we propose a partitioning strategy of the bath modes for multiple parallel chains to reduce primary mode coupling strengths and also the number of the modes directly coupled to the system operator. 
This is shown in Figs.~\ref{fig:bathmodel}{\bf C} and {\bf D} respectively.  
We found that the coupling strength of the primary modes, which are directly coupled to the system, can be reduced as we increase the number of the chains; at the same time, we can also shorten the lengths of each chain. In addition to the fabrication and implementation benefits for open quantum simulators using quantum hardware, these methods are also potentially applicable to simulations in classical computers. In this case, perturbative methods may be employed to simulate these chain models with reduced system-bath coupling~\cite{Burghardt2012,Chenel2014}.

A recurrence equation derived by Bulla \emph{et al.}~\cite{Bulla2005} has been used in the renormalization group approaches to construct the 1-D Wilson chain (Fig.~\ref{fig:bathmodel}{\bf B}). This recurrence relation, however, potentially shows numerical instabilities~\cite{Vojta_2005,Bulla2005,Bulla_2008}. Recently, Chin \emph{et al.}~\cite{Chin_2010} developed an exact mapping approach for a continuous SDF using orthogonal polynomials without discretizing the SDF. However, this approach may pose challenges applicable to arbitrary structured SDFs. In many applications of chemistry and biology, structured SDFs appear when atomistic details are involved in open quantum dynamics, as already mentioned in the introduction. Therefore, the SDFs may not be well approximated as simple analytic functions such as Ohmic spectral functions. 

In this paper, we test a generalized linear algebraic transformation approach for any given discrete SDFs, where a transformation on multiple parallel chains is involved, as shown in Fig.~\ref{fig:bathmodel}{\bf C}. 
With the multiple chain-bath transformation described in the following sections, complex open quantum systems, such as photosynthetic antennae, can be studied practically via quantum simulators.

In the next sections we present the model Hamiltonian and the linear algebraic bath transformation. As an example, a two-oscillator bath transformation is presented analytically. This example shares many features of our general scheme of bath partitioning. Numerical stability of the bath transformation methods is discussed in the result section and the results are compared with Bulla's transformation approach~\cite{Bulla2005}. Then, we propose a way to partition the bath modes into multiple parallel chains to reduce the system-bath coupling strengths. We apply the proposed leaping partitioning (LP) strategy to a structured spectral density of the chlorosome~\cite{Fujita2012,Fujita2014}, as an example. The numerical result is compared with a 'standard' sequential partitioning (SP) scheme.



\section{Chain bath transformation}
As mentioned in the introduction, in the theory of open quantum systems, the system-bath Hamiltonian $\hat{H}$ is composed of three parts, namely,
\begin{equation}
\hat{H}=\hat{H}_{\mathrm{S}}+\hat{H}_{\mathrm{SB}}+\hat{H}_{\mathrm{B}} \ ,
\end{equation}
where $\hat{H}_{\mathrm{S}}$ is the system Hamiltonian. The phonon bath $\hat{H}_{\mathrm{B}}$ is approximated as a set of non-interacting harmonic oscillators. 
The coupling term $\hat{H}_{\mathrm{SB}}$ between the system and bath is almost universally treated as a bilinear coupling. More precisely, we write $\hat{H}_{\mathrm{SB}}+\hat{H}_{\mathrm{B}}$ in a compact form~\cite{Guo:2012} as follows,
\begin{align}
&\hat{H}_{\mathrm{SB}}+\hat{H}_{\mathrm{B}}=\sum_{n}
\begin{pmatrix}
\hat{L}_{n} & \vec{\hat{a}}_{n}^{\dagger}
\end{pmatrix}
\matnew{\Gamma}_{n}
\begin{pmatrix}
\hat{L}_{n} \\ \vec{\hat{a}}_{n}
\end{pmatrix} \label{eq:starbath} , \\
&\matnew{\Gamma}_{n}=\begin{pmatrix}
0&\vec{\kappa}^{\transpose}_{n}\\
\vec{\kappa}_{n}&\matnew{\Omega}_{n}
\end{pmatrix},\label{eq:Omega0}
\end{align}
where each  creation (annihilation) operator vector $\vec{\hat{a}}_{n}^\dagger\, \,  (\vec{\hat{a}}_{n})$ of oscillators for the site $n$ are coupled to the operator $\hat{L}_{n}$  that acts on the system. 
Lower case bold and capital bold fonts are used for a column vector and a matrix, respectively. 
The bosonic operators satisfy a commutation relation, $[\hat{a}_{n;i},\hat{a}_{m;j}^{\dagger}]=\delta_{nm}\delta_{ij}$ where $\vec{\hat{a}}_{n}=(\hat{a}_{n;1},\ldots,\hat{a}_{n;N})^{\transpose}$. Here $\matnew{\Omega}_{n}$ is a diagonal matrix, which has the harmonic frequencies as the elements, \emph{i.e.} $\matnew{\Omega}_{n}=\diag(\omega_{n;1},\ldots,\omega_{n;N})$. 
$\vec{\kappa}_{n}$ is the system-bath coupling strength vector.
Accordingly, the SDF $J_{n}(\omega)$ is defined~\cite{Valleau_2012} as,
\begin{equation}
J_{n}(\omega)=\pi\sum_{j}\kappa_{n;j}^{2}\,\delta(\omega-\omega_{n;j}) \ .
\end{equation}
The non-interacting bath in Eq.~\ref{eq:starbath} is the star-bath model (Fig.~\ref{fig:bathmodel}{\bf A}), where the independent harmonic oscillators are all coupled directly to the system.


\subsection{Linear algebraic bath transformation}
With a suitable choice of unitary transformation on the bath oscillators, one can turn a star-bath into a multiple-chain bath. The multiple-chain bath has a few primary bath oscillators and the remaining oscillators (secondary bath modes) are coupled to the primary bath modes in a chain as depicted in Fig.~\ref{fig:bathmodel}{\bf C}. A mixture of star and chain models is also possible as shown in Fig.~\ref{fig:bathmodel}{\bf D}.
Burghardt \emph{et al.}~\cite{Hughes2009,Burghardt2012,Chenel2014} exploited the latter model to develop a perturbative truncated bath model, which approximates the terminal star-coupled yellow oscillators in Fig.~\ref{fig:bathmodel}{\bf D} as Markovian baths.

The bath transformation from the star model (Eq.~\ref{eq:starbath}) to the 1-D Wilson chain (Fig.~\ref{fig:bathmodel}{\bf B}) can be simply obtained by a unitary transformation of the matrix $\matnew{\Gamma}_{n}$ that keeps the system operators unchanged. We introduce, here, an arbitrary unitary transformation ($\matnew{U}_{n}\matnew{U}_{n}^{\dagger}=\matnew{I}$) satisfying the following conditions;
\begin{align}
&\hat{H}_{\mathrm{SB}}+\hat{H}_{\mathrm{B}}=\sum_{n}
\begin{pmatrix}
\hat{L}_{n} & \vec{\hat{b}}_{n}^{\dagger}
\end{pmatrix}
\matnew{\tilde{\Gamma}}_{n}
\begin{pmatrix}
\hat{L}_{n} \\ \vec{\hat{b}}_{n}
\end{pmatrix}, \label{eq:chainbath}\\
&\matnew{\tilde{\Gamma}}_{n}=\begin{pmatrix}
0&\vec{\tilde{\kappa}}^{\transpose}_{n}\\
\vec{\tilde{\kappa}}_{n}&\matnew{\tilde{\Omega}}_{n}
\end{pmatrix}
=\begin{pmatrix}
1&\vec{0}^{\transpose}\\
\vec{0}&\matnew{U}_{n}^{\dagger}
\end{pmatrix}
\matnew{\Gamma}_{n}
\begin{pmatrix}
1&\vec{0}^{\transpose}\\
\vec{0}&\matnew{U}_{n}
\end{pmatrix}
, \label{eq:GSHU}
\end{align}
where $\vec{\hat{b}}_{n}=\matnew{U}_{n}^{\dagger}\vec{\hat{a}}_{n}$. 
The first column of $\matnew{U}_{n}$ is $\vert\vert \vec{\kappa}_{n} \vert\vert^{-1}_{2}\vec{\kappa}_{n}$  and the remaining columns are constructed using the Gram-Schmidt process with random vectors (or unit vectors)~\cite{Golub:1996}. As a result, $\matnew{\tilde{\Gamma}}_{n}$ is a dense symmetric matrix and \mbox{$\vec{\tilde{\kappa}}_{n}=(\tilde{\kappa}_{n;1},0,\ldots,0)^{\transpose}$} is the new system-bath coupling strength vector. (see Appendix A for the details and down panels of Fig.~\ref{fig:bathmodel} for the structures of $\matnew{\tilde{\Gamma}}_{n}$.)

Now we have new sets of interacting harmonic oscillators while the system operators remain unchanged. 
The tridiagonalization of $\matnew{\Gamma}_{n}$ in Eq.~\ref{eq:starbath} for the Wilson chain (Fig.~\ref{fig:bathmodel}{\bf B}) can be performed numerically by Householder or Lanczos procedures~\cite{Golub:1996}. Alternatively, we use here tridiagonalization of $\matnew{\tilde{\Omega}}_{n}$, \emph{i.e.} $\matnew{\tilde{\Omega}}_{n}=\matnew{T}\matnew{\Xi}\matnew{T}^{\dagger}$, with a Hessenberg reduction of a symmetric matrix~\cite{Golub:1996}.
We call the later transformation method as Gram-Schmidt-Hessenberg (GSH).
The Hessenberg reduction of a symmetric matrix produces a tridiagonal matrix and then the numerical procedures for the reduction, such as, Householder, Lanczos and Gauss transform, can be applied. 
These numerical algorithms are standard numerical linear algebraic techniques, see \emph{e.g.} Ref.~\cite{Golub:1996}.


\subsection{Multiple chain transformation}


In this subsection we explain the multiple parallel chain transformation that is depicted in Fig.~\ref{fig:bathmodel}{\bf C}.
We introduce a unitary transformation,
\begin{equation}
\matnew{\tilde{U}}_{n}=\matnew{P}_{n}\matnew{U}_{n} \ ,
\end{equation}
that additionally rearranges (using a permutation matrix $\matnew{P}_{n}$)  the non-interacting bath oscillators 
as multiple groups of several interacting oscillators, \emph{i.e.}
\mbox{$\vec{\hat{b}}_{n}=\matnew{\tilde{U}}_{n}^{\dagger}\vec{\hat{a}}_{n}$}.
The unitary transformation matrix  $\matnew{U}_{n}$ is block diagonal and does not allow the interaction between oscillators from different groups (an example of the rearrangement is given in Appendix B.).  
We also define the following relations for the unitary transformation;
\begin{equation}
\matnew{\tilde{\Omega}}_{n}=
\matnew{\tilde{U}}_{n}^{\dagger}
\matnew{\Omega}_{n}\matnew{\tilde{U}}_{n}
\quad 
\mathrm{and}
\quad
\vec{\tilde{\kappa}}_{n}=
\matnew{\tilde{U}}_{n}^{\dagger}\vec{\kappa}_{n} \ .
\end{equation}

The primary modes, which are directly coupled to the system operators, are defined as collective oscillator modes by choosing the first column of the $l$-th subblock $\matnew{U}_{n}^{(l)}$ to be $\vert\vert \vec{g}^{(l)}_{n} \vert\vert^{-1}_{2}\vec{g}^{(l)}_{n}$. The normalized vector corresponds to the rearranged coupling strength vector of
\begin{equation}
\vec{g}_{n}=\matnew{P}_{n}^{\dagger}\vec{\kappa}_{n}=
\begin{pmatrix}
\vec{g}^{(1)}_{n}\\
\vdots \\
\vec{g}^{(N_{eff})}_{n}
\end{pmatrix},
\end{equation}
with $N_{eff}$ being the number of subblocks (or the number of group of oscillators). 
The chain-bath model can be obtained via tridiagonalization of $l$-th subblock
\begin{equation}
\matnew{\tilde{\Omega}}_{n}^{(l)}=\matnew{T}^{(l)}\matnew{\Xi}^{(l)}\matnew{T}^{(l)\dagger} \ 
\end{equation}
with applying the Hessenberg transform~\cite{Golub:1996} via the Householder procedure. 
$\matnew{\Xi}^{(l)}$ is a tridiagonal matrix that defines the frequencies (diagonal elements) of the transformed bath modes and coupling strengths (off-diagonal elements) between the oscillators in the chain model. $\matnew{T}^{(l)}$ is a Hessenberg unitary transform matrix that makes no transformation to the primary bath mode such that the first column of the matrix is $(1,0,\ldots,0)^{\transpose}$.
%
The resulting transformed  bath coupling vector $\matnew{T}^{(l)}\vec{\tilde{\kappa}}_{n}^{(l)}$ of the $l$-th subblock has only a single non-zero first element, which corresponds to the primary mode coupling strength. In Appendix C, we provide a MATLAB~\cite{MATLAB:2013} code for the GSH with the LP scheme. 

The alternative numerical transformation from the star-bath model to the chain-bath model can be obtained by Bulla's recursion method~\cite{Bulla2003,Bulla2005}. The two methods will be compared numerically later.


\section{Results and discussion}


\subsection{Numerical stability of the transformations}
\begin{figure}[t]
\begin{center}
\resizebox{\linewidth}{!}{\includegraphics{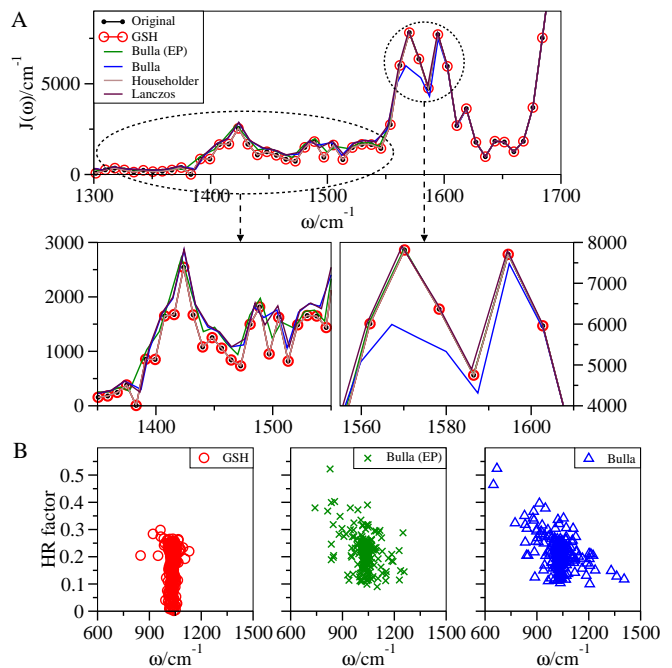}}
\caption{Reconstructed spectral densities of the chlorosome from a chain bath. {\bf A}. Reconstructed spectral densities of chlorosome from different numerical transformations are compared with the original one. {\bf B}. Huang-Rhys (HR) factors of the secondary bath oscillators  with nearest-neighbor couplings in a single chain model. Two different unitary mapping methods are compared. Red circles are calculated by the Gram-Schmidt-Hessenberg (GSH) transform. Green crosses and blue triangles are obtained from Bulla's method~\cite{Bulla2005} with and without the extended precision (EP), respectively.
}\label{fig:hrmethods}
\end{center}
\end{figure}

To test the numerical stability of the 1-D Wilson chain transformation methods, we perform back transformations from the chain-bath Hamiltonian to the star-bath Hamiltonian (Eq.~\ref{eq:starbath})  by a straightforward diagonalization of  $\matnew{\tilde{\Omega}}_{n}$. The structured SDFs are reconstructed by the back transformation and then compared with the original SDF of the chlorosome, as an example system.
The chlorosome is a giant light harvesting antenna complex of green sulfur bacteria~\cite{blankenshipbook}. The excitation energy is transferred within the antenna via the fluctuating environment. Various models were developed to study the system from the open quantum dynamics perspective~\cite{Fujita2012,Fujita2014,Huh:2014}. Here we use the SDF of the chlorosome that some of us~\cite{Fujita2012} obtained via quantum mechanics/molecular mechanics (QM/MM) calculations that contain 253 peaks corresponding to the quantum bath oscillators.
In Fig.~\ref{fig:hrmethods}{\bf A}, only the peaks in 1300--1700 cm$^{-1}$ are shown for clarity.
The original SDF is plotted as a black line and filled black circles. Bulla's method (blue line) suffers from numerical instability as the iteration increases. Therefore, we also test extended precision (EP; 100 digits) with Bulla's method (green line). The GSH curve is shown as a red line and unfilled red circles. Householder and Lanczos transformations of $\matnew{\Gamma}_{n}$ in Eq.~\ref{eq:Omega0} are plotted in brown and purple lines, respectively. As expected, Bulla's original method generates a curve that deviates significantly from the original one, especially around 1550--1600 cm$^{-1}$. The EP improves the result but is still in disagreement with the original. The Lanczos curve seems to agree well with the original but the discrete data points do not match with the original data points in the frequency domain and it produces negative frequencies that are not shown in the figure. The GSH and Householder, on the other hand, can reproduce the original SDF with high accuracy. Both methods are based on the Householder procedure, which has an unconditional stability~\cite{Businger1969}. 
%

Fig.~\ref{fig:hrmethods}{\bf B} indicates the Huang-Rhys (HR) factors of secondary bath oscillators with nearest-neighbor couplings in a chain, that are obtained from different methods.
The HR factor $\chi_{j}$ of a harmonic oscillator with frequency $\omega_{j}$ is a normalized coupling strength given by $\kappa_{j}=\omega_{j}\sqrt{\chi_{j}}$. 
For the secondary modes, HR factors are defined with the frequencies of the oscillators and the coupling strengths with the nearest neighbors in the chain. As one compares the results from Bulla's method (green cross) and Bulla's method with EP (blue triangle), they are significantly different from each other.
This shows that Bulla's recursion relation is numerically unstable.
Comparing the Bulla's and GSH methods, one can see the Bulla's method (green cross) produces larger HR factors than the GSH method (red circle).
The GSH bath transformation can produce oscillators with frequencies distributed a narrower band. The reason for this difference is that the unitary transformation for the chain model is not unique. Bulla's method does not allow the secondary bath modes to have negative interactions while the GSH has no such a constraint. Since the unitary matrix for the GSH transform is constructed via orthogonalization of the random vectors, the signs of the interactions in a single chain can vary in each transformation, but their magnitudes are invariant.


\subsection{Multiple chain bath model}
\begin{figure}[t]
\begin{center}
\includegraphics[width=0.85\linewidth,angle=-90]{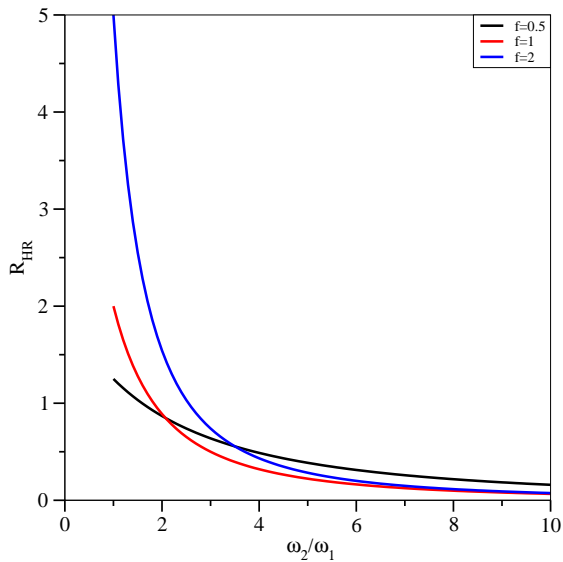}
\caption{Normalized Huang-Rhys (HR) factor $R_{\mathrm{HR}}$ for the  example of bath with two oscillator modes. With selected fixed $f$ values, the normalized HR factors are evaluated at varying the frequency ratio ($\omega_{2}/\omega_{1}$).
}\label{fig:twodimhr}
\end{center}
\end{figure}

\begin{figure}[t]
\begin{center}
\resizebox{\linewidth}{!}{\includegraphics{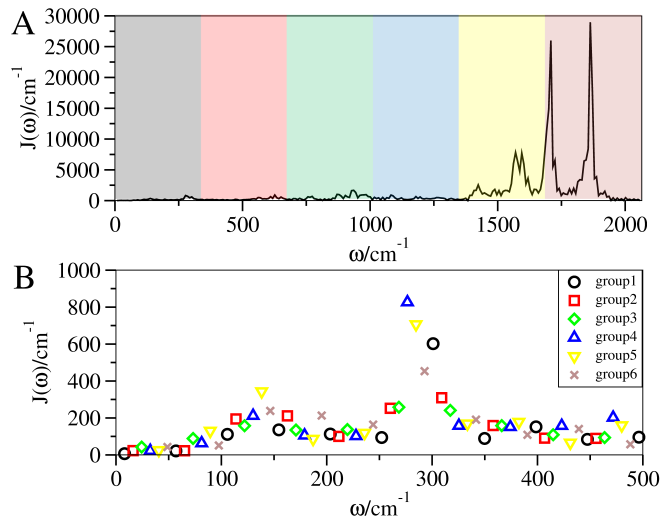}}
\caption{Partitioning schemes of the spectral density function (SDF). {\bf A}. The sequential partition  of the spectral density of chlorosome is depicted. Group of oscillators are indicated with 6 different colors. {\bf B}. The leaping partition of the spectral density is depicted. The peaks in the spectral density of the chlorosome~\cite{Fujita2012,Fujita2014} are partitioned into 6 groups. Only the peaks below 500 cm$^{-1}$ are shown for the clarity.
}\label{fig:chlspd}
\end{center}
\end{figure}

In this subsection, we apply the GSH transformation method to a couple of illustrative examples.
First, to get some insights into the weakly-coupled multiple chain model, we employ this method analytically  to transform  a bath with two oscillator modes. 
%
Then, we continue with a multidimensional example of a chlorosome. 
%

In the chain bath model for two oscillators, the mixing  of the two modes with frequencies $\omega_{1}\le\omega_{2}$ leads to a HR factor for the primary mode $\chi_{\mathrm{c}}$:
\begin{align}
&\chi_{\mathrm{c}}=R_{\mathrm{HR}}\chi_{1},\\
&R_{\mathrm{HR}}=\frac{(1+f^{2})^{3}}{(1+\tfrac{\omega_{2}}{\omega_{1}}f^{2})^{2}},\label{eq:RHR}
\end{align}
where $f=\kappa_{2}/\kappa_{1}$ and $\chi_{1}=\kappa_{1}^{2}/\omega_{1}^{2}$. In Fig.~\ref{fig:twodimhr}, the normalized HR factor $R_{\mathrm{HR}}$ of Eq.~\ref{eq:RHR} is plotted while varying the frequency ratio ($\omega_{2}/\omega_{1}$) at fixed coupling strength ratios $f$, 0.5, 1 and 2. 
For fixed $f$, the values of $R_{\mathrm{HR}}$ decrease as the frequency ratio increases. 
$f$ determines the slopes of the curves. Larger $f$ makes the curve decrease faster as $\omega_{2}/\omega_{1}$ increases.
When the oscillators have similar frequencies $\omega_{2}/\omega_{1}\simeq 1$, $R_{\mathrm{HR}}$ is larger than 1, which makes the chain bath couplings stronger than the star-bath model. However, as the frequency ratio increases $R_{\mathrm{HR}}$ drops down and it can become arbitrarily small as the frequency difference increases. 
This gives a hint how to mix the oscillator modes and reduce the coupling strength of the primary modes by forming weakly coupled multiple chains. The oscillators, which have large frequency differences, should be mixed to make the weakly coupled multiple chain bath models.

Next, we apply the GSH method to the example of chlorosome. 
The SP approach divides the bath oscillators into a sequence of groups of oscillators, as illustrated in Fig.~\ref{fig:chlspd}{\bf A} using a different color for each group.
Fig.~\ref{fig:chlspd}{\bf B} represents the LP scheme, where only the peaks below 500 cm$^{-1}$ are shown for the clarity.
As indicated here, the oscillators are partitioned into 6 groups to use the LP scheme. In the LP scheme,  the elements of the $k$-th group are composed of $k+N_{\mathrm{eff}}j$-th ($\le N$) modes, where $N_{\mathrm{eff}}$ is the number of groups, $j$ is an integer ($\ge 0$) and $N$ is the total number of oscillators. For example, when one has 10 modes and 3 groups to partition, 
the SP approach groups modes of (1,2,3), (4,5,6) and (7,8,9,10) as group 1, 2 and 3, respectively. In the other hand, the LP strategy does the partition into (1,4,7,10), (2,5,8) and (3,6,9) for group 1, 2 and 3, respectively. 

\begin{figure}[t]
\begin{center}
\resizebox{\linewidth}{!}{\includegraphics[angle=-90]{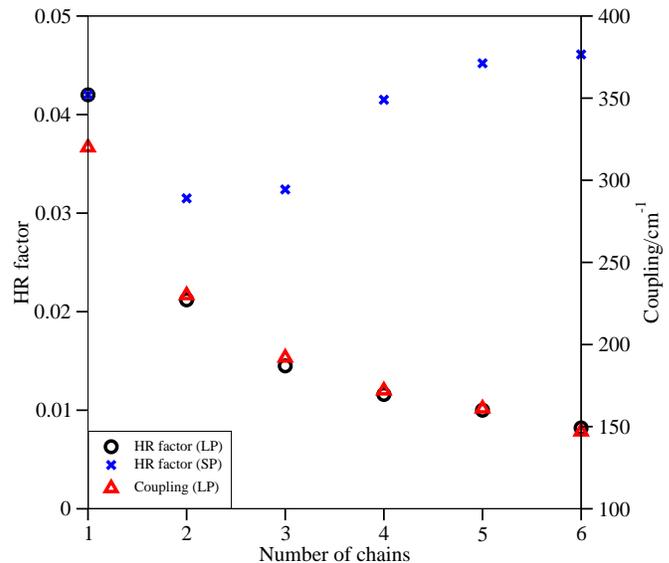}}
\caption{Maximum Huang-Rhys (HR) factor of primary modes of the chlorosome. The maximum HR factor (black circles) of primary modes and the corresponding coupling strengths (red triangles) are plotted as increasing the number of chains for the leaping partitioning (LP) scheme. Blue cross is used for the maximum HR factors from the sequential partitioning (SP) strategy.
}\label{fig:numofchains}
\end{center}
\end{figure}

In Fig.~\ref{fig:numofchains}, the maximum HR factor and the corresponding coupling strength of the primary modes are plotted by varying the number of chains from 1 to 6. 
The LP and SP schemes are used for this calculation. 
The maximum HR factor of the star-bath model is 0.0315. 
The single chain model has even larger HR factor of the primary mode than the star-bath model for both partitioning schemes.
The maximum HR factors from the SP scheme (blue cross) do not decrease as the number of chains increases.
The figure shows, however, that the maximum HR factor (black circle) and the corresponding coupling strength (red triangle) decrease as the number of chains increases for the LP scheme. The maximum HR factor from the LP scheme with 6 chains is below 0.01, which corresponds to 10\% of the corresponding harmonic frequency. Therefore 6 chains make the chain model suitable for implementation on quantum simulators, since a quantum simulator can have only a few parallel chains and, in addition, the primary mode HR factor is limited. In principle, the values can be reduced further as long as mixing modes is still possible.

\begin{figure}[t]
\begin{center}
\includegraphics[width=0.85\linewidth,angle=-90]{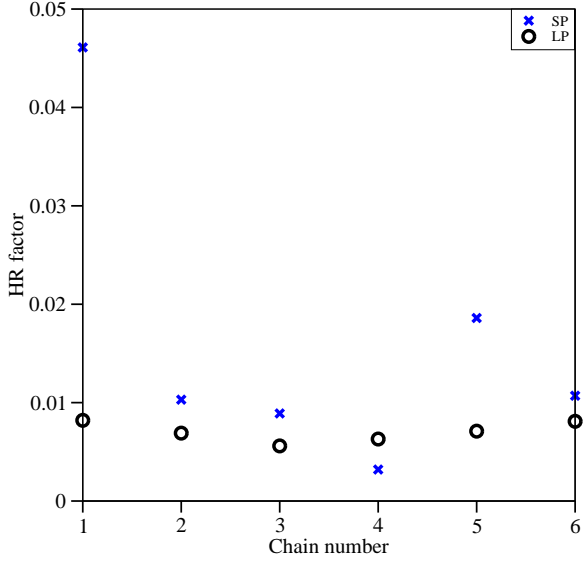}
\caption{Huang-Rhys (HR) factor of each chain of the chlorosome in different partition schemes. The bath oscillators of chlorosome are transformed to 6 chain bath model with the two different grouping strategies. The sequential partitioning (SP; blue cross) and leaping partitioning (LP; black circle) strategies are compared. Lower HR factors lead to more feasible quantum simulator schemes.
}\label{fig:parchains}
\end{center}
\end{figure}

The LP and SP schemes are further compared in Fig.~\ref{fig:parchains}.
This plot shows the HR factor of the primary modes of each chain. 
The results  are indicated in blue crosses and black circles for the SP and LP schemes, respectively. 
As evidenced by the figure, the LP scheme gives all values below 0.01 while the SP strategy produce bigger values and the maximum value is more than 4 times larger than the LP values. Another important aspect of the LP scheme is that all of the primary HR factors have nearly similar values while the results of the SP scheme deviate largely from each other.

\section{Conclusion and outlook}
In this paper we show that a multiple chain-bath model, in combination with the leaping bath partitioning scheme, may lead to a practical implementation of quantum simulators for complex open quantum dynamics.
We have shown that the multiple chain-bath model can be employed for the realization of quantum simulators for open quantum systems or for numerical studies in classical computers. Furthermore, the leaping partition scheme can reduce the primary mode coupling strength almost homogeneously for all parallel chains. The reason is that the mixing of oscillators with large frequency differences can result in small HR factors. The two-oscillator model was presented with an analytic expression for the chain transformation and provides a hint for the bath partitioning scheme, \emph{i.e.} leaping partition. We also tested the unitary transformation algorithm that exploits 
the GSH transformation, and compared the results with the values from Bulla's recursion method~\cite{Bulla2005}. The GSH transformation can produce smaller HR factors of the secondary bath modes with oscillators being in a narrower band. The numerical stability of Bulla's method was discussed and the GSH method was shown to be numerically stable.

Our bath transformation method could be useful for the perturbative approaches as well, because of the resulting weak system-bath couplings. The effective spectral density can be obtained based on the chain-model transformation. It can also be used for the reduced density matrix methods~\cite{Burghardt2012,Chenel2014} for the simulation of non-Markovian dynamics.
The effective bath approach with the parallel chain-bath model can also be useful for the HEOM method. The effective bath spectral density can reduce the number of Drude-Lorentzian peaks, then the HEOM method can handle larger systems~\cite{Mostame:2014}. Further work in this direction will be conducted.

\section*{Acknowledgements}
We acknowledge Jarrod McClean, Florian Schroeder, Dr. Christoph Kreisbeck and Dr. Semion K. Saikin for discussions. J. H., T. F. and A. A.-G. acknowledge support from the Center for Excitonics, an Energy Frontier Research Center funded by the US Department of Energy, Office of Science and Office of Basic Energy Sciences under award DE-SC0001088.
J. H. and S. M. and A. A.-G. also acknowledge Defense Threat Reduction Agency grant HDTRA1-10-1-0046 and the Air Force Office of Scientific Research grant FA9550-12-1-0046.
Further, A. A.-G. is grateful for the support from Defense Advanced Research Projects Agency grant N66001-10-1-4063, and the Corning Foundation for their generous support. 
M.-H. Y. acknowledges the support by the National Basic Research Program
of China Grant 2011CBA00300, 2011CBA00301, the National Natural
Science Foundation of China Grant 61033001, 61361136003, and the Youth
1000-talent program.
%
\appendix
\section{GSH algorithm}
We explain here in details how to construct the unitary transform matrix 
$\matnew{U}_{n}$ of GSH transformation in Eq.~\ref{eq:GSHU} for $\matnew{\tilde{\Omega}}_{n}$. A MATLAB~\cite{MATLAB:2013} script is given in Appendix C. \\

1. Set the first column of $\matnew{U}_{n}$ to be the normalized coupling strength vector $\vert\vert \vec{\kappa}_{n} \vert\vert^{-1}_{2}\vec{\kappa}_{n}$.\\
\indent 2. Assign random vectors to the remaining columns (from 2 to $N$). We use a normal distribution with the first column to be a mean vector for the random vectors.\\
\indent 3. Perform the Gram-Schmidt orthogonalization to~$\matnew{U}_{n}$.\\
\indent 4. Compute $\matnew{U}_{n}^{\dagger}\matnew{\Omega}_{n}\matnew{U}_{n}$ for $\matnew{\tilde{\Omega}}_{n}$.

\section{Example of permutation matrix}
Here we represent a permutation matrix $\matnew{P}^{\dagger}$, as an example. 
It permutes a vector of length 4 ($\vec{v}$) to arrange odd and even elements sequentially,
\begin{align}
\matnew{P}^{\dagger}\vec{v}=
\begin{pmatrix}
1 & 0 & 0 & 0\\
0 & 0 & 1 & 0\\
0 & 1 & 0 & 0\\
0 & 0 & 0 & 1
\end{pmatrix}
\begin{pmatrix}
v_{1}\\
v_{2}\\
v_{3}\\
v_{4}
\end{pmatrix}
=
\begin{pmatrix}
v_{1}\\
v_{3}\\
v_{2}\\
v_{4}
\end{pmatrix} \ .
\end{align}

\section{LP scheme with GSH}
We provide here a  MATLAB~\cite{MATLAB:2013} script for the LP scheme with the GSH procedure. The script produces a single chain for a subblock, which is assigned by the user. 
Multiple chains can be obtained by using this script for each subblock. 
The details of the script can be found in the comments of the script,  which are indicated with \%. 
\begin{lstlisting}
function [HRF,Dn,Xi]=LP(SDF,Sub,Neff)
%-Inputs-
%SDF: Spectral density function = [frequency spectral-density]
%Sub: Subblock of interest (starting index in the full SDF)
%Neff: Number of subblocks (index step size for LP scheme)

%-Outputs-
%HRF: HR factor of the primary mode of the subblock
%Dn: norm of the coupling vector (equivelent to the coupling strength of the primary mode)
%Xi: Tridiagonal coupling bath coupling matrix
%    Eq. (10), OmegaTilde=T*Xi*T', T*T'=I

%Scale SDF with 1/pi
SDF(:,2)=SDF(:,2)/pi;
%dimension of the SDF
dim=max(size(SDF));

%LP scheme for the subgroup
SDF=spd(Sub:Neff:dim,:);
%new dimension
dim=max(size(SDF));

%t1: first column of U, U*U'=I
t1=sqrt(SDF(:,2));
Dn=norm(t1);
t1=t1/Dn;

%Random vectors for the Gram-Schmidt orthogonalization
%are assigned to columns from 2 to N. 
U0=randn(dim,dim-1);
U0=[t1 U0];
for k=2:dim
U0(:,k)=U0(:,k)+t1;
end
%Gram-Schmidt orthogonalization, user should provide the GS routine. 
U=GS(U0);
%OmegaTilde in Eq. 8
OmegaTilde=U'*diag(SDF(:,1))*U;
%Hessenberg transformation
[T,Xi]=hess(OmegaTilde);
HRF=Dn^2/Xi(1,1)^2;

end
\end{lstlisting}
%
\section*{References}

\begin{thebibliography}{75}%
\makeatletter
\providecommand \@ifxundefined [1]{%
 \@ifx{#1\undefined}
}%
\providecommand \@ifnum [1]{%
 \ifnum #1\expandafter \@firstoftwo
 \else \expandafter \@secondoftwo
 \fi
}%
\providecommand \@ifx [1]{%
 \ifx #1\expandafter \@firstoftwo
 \else \expandafter \@secondoftwo
 \fi
}%
\providecommand \natexlab [1]{#1}%
\providecommand \enquote  [1]{``#1''}%
\providecommand \bibnamefont  [1]{#1}%
\providecommand \bibfnamefont [1]{#1}%
\providecommand \citenamefont [1]{#1}%
\providecommand \href@noop [0]{\@secondoftwo}%
\providecommand \href [0]{\begingroup \@sanitize@url \@href}%
\providecommand \@href[1]{\@@startlink{#1}\@@href}%
\providecommand \@@href[1]{\endgroup#1\@@endlink}%
\providecommand \@sanitize@url [0]{\catcode `\\12\catcode `\$12\catcode
  `\&12\catcode `\#12\catcode `\^12\catcode `\_12\catcode `\%12\relax}%
\providecommand \@@startlink[1]{}%
\providecommand \@@endlink[0]{}%
\providecommand \url  [0]{\begingroup\@sanitize@url \@url }%
\providecommand \@url [1]{\endgroup\@href {#1}{\urlprefix }}%
\providecommand \urlprefix  [0]{URL }%
\providecommand \Eprint [0]{\href }%
\providecommand \doibase [0]{http://dx.doi.org/}%
\providecommand \selectlanguage [0]{\@gobble}%
\providecommand \bibinfo  [0]{\@secondoftwo}%
\providecommand \bibfield  [0]{\@secondoftwo}%
\providecommand \translation [1]{[#1]}%
\providecommand \BibitemOpen [0]{}%
\providecommand \bibitemStop [0]{}%
\providecommand \bibitemNoStop [0]{.\EOS\space}%
\providecommand \EOS [0]{\spacefactor3000\relax}%
\providecommand \BibitemShut  [1]{\csname bibitem#1\endcsname}%
\let\auto@bib@innerbib\@empty
\bibitem [{\citenamefont {Breuer}\ and\ \citenamefont
  {Petruccione}(2006)}]{BRE02}%
  \BibitemOpen
  \bibfield  {author} {\bibinfo {author} {\bibfnamefont {H.~P.}\ \bibnamefont
  {Breuer}}\ and\ \bibinfo {author} {\bibfnamefont {F.}~\bibnamefont
  {Petruccione}},\ }\href@noop {} {\emph {\bibinfo {title} {{The theory of open
  quantum systems}}}}\ (\bibinfo  {publisher} {Oxford University Press},\
  \bibinfo {address} {New York},\ \bibinfo {year} {2006})\BibitemShut {NoStop}%
\bibitem [{\citenamefont {May}\ and\ \citenamefont
  {K\"uhn}(2011)}]{May:1320614}%
  \BibitemOpen
  \bibfield  {author} {\bibinfo {author} {\bibfnamefont {V.}~\bibnamefont
  {May}}\ and\ \bibinfo {author} {\bibfnamefont {O.}~\bibnamefont {K\"uhn}},\
  }\href@noop {} {\emph {\bibinfo {title} {{Charge and Energy Transfer Dynamics
  in Molecular Systems; 3rd ed.}}}}\ (\bibinfo  {publisher} {Wiley-VCH},\
  \bibinfo {address} {Weinheim},\ \bibinfo {year} {2011})\BibitemShut {NoStop}%
\bibitem [{\citenamefont {van Amerongen}\ \emph {et~al.}(2000)\citenamefont
  {van Amerongen}, \citenamefont {Valkunas},\ and\ \citenamefont {van
  Grondelle}}]{grondelle00}%
  \BibitemOpen
  \bibfield  {author} {\bibinfo {author} {\bibfnamefont {H.}~\bibnamefont {van
  Amerongen}}, \bibinfo {author} {\bibfnamefont {L.}~\bibnamefont {Valkunas}},
  \ and\ \bibinfo {author} {\bibfnamefont {R.}~\bibnamefont {van Grondelle}},\
  }\href@noop {} {\emph {\bibinfo {title} {{Photosynthetic excitons}}}}\
  (\bibinfo  {publisher} {World Scientific},\ \bibinfo {address} {Singapore},\
  \bibinfo {year} {2000})\BibitemShut {NoStop}%
\bibitem [{\citenamefont {Bulla}\ \emph {et~al.}(2003)\citenamefont {Bulla},
  \citenamefont {Tong},\ and\ \citenamefont {Vojta}}]{Bulla2003}%
  \BibitemOpen
  \bibfield  {author} {\bibinfo {author} {\bibfnamefont {R.}~\bibnamefont
  {Bulla}}, \bibinfo {author} {\bibfnamefont {N.-H.}\ \bibnamefont {Tong}}, \
  and\ \bibinfo {author} {\bibfnamefont {M.}~\bibnamefont {Vojta}},\
  }\href@noop {} {\bibfield  {journal} {\bibinfo  {journal} {Phys. Rev. Lett.}\
  }\textbf {\bibinfo {volume} {91}},\ \bibinfo {pages} {170601} (\bibinfo
  {year} {2003})}\BibitemShut {NoStop}%
\bibitem [{\citenamefont {Bulla}\ \emph {et~al.}(2005)\citenamefont {Bulla},
  \citenamefont {Lee}, \citenamefont {Tong},\ and\ \citenamefont
  {Vojta}}]{Bulla2005}%
  \BibitemOpen
  \bibfield  {author} {\bibinfo {author} {\bibfnamefont {R.}~\bibnamefont
  {Bulla}}, \bibinfo {author} {\bibfnamefont {H.-J.}\ \bibnamefont {Lee}},
  \bibinfo {author} {\bibfnamefont {N.-H.}\ \bibnamefont {Tong}}, \ and\
  \bibinfo {author} {\bibfnamefont {M.}~\bibnamefont {Vojta}},\ }\href@noop {}
  {\bibfield  {journal} {\bibinfo  {journal} {Phys. Rev. B}\ }\textbf {\bibinfo
  {volume} {71}},\ \bibinfo {pages} {045122} (\bibinfo {year}
  {2005})}\BibitemShut {NoStop}%
\bibitem [{\citenamefont {Huh}\ \emph {et~al.}(2014)\citenamefont {Huh},
  \citenamefont {Saikin}, \citenamefont {Brookes}, \citenamefont {Valleau},
  \citenamefont {Fujita},\ and\ \citenamefont {Aspuru-Guzik}}]{Huh:2014}%
  \BibitemOpen
  \bibfield  {author} {\bibinfo {author} {\bibfnamefont {J.}~\bibnamefont
  {Huh}}, \bibinfo {author} {\bibfnamefont {S.~K.}\ \bibnamefont {Saikin}},
  \bibinfo {author} {\bibfnamefont {J.~C.}\ \bibnamefont {Brookes}}, \bibinfo
  {author} {\bibfnamefont {S.}~\bibnamefont {Valleau}}, \bibinfo {author}
  {\bibfnamefont {T.}~\bibnamefont {Fujita}}, \ and\ \bibinfo {author}
  {\bibfnamefont {A.}~\bibnamefont {Aspuru-Guzik}},\ }\href@noop {} {\bibfield
  {journal} {\bibinfo  {journal} {J. Am. Chem. Soc.}\ }\textbf {\bibinfo
  {volume} {136}},\ \bibinfo {pages} {2048} (\bibinfo {year}
  {2014})}\BibitemShut {NoStop}%
\bibitem [{\citenamefont {Ishizaki}\ and\ \citenamefont
  {Tanimura}(2005)}]{Ishizaki2005}%
  \BibitemOpen
  \bibfield  {author} {\bibinfo {author} {\bibfnamefont {A.}~\bibnamefont
  {Ishizaki}}\ and\ \bibinfo {author} {\bibfnamefont {Y.~J.}\ \bibnamefont
  {Tanimura}},\ }\href@noop {} {\bibfield  {journal} {\bibinfo  {journal} {J.
  Phys. Soc. Jpn.}\ }\textbf {\bibinfo {volume} {74}},\ \bibinfo {pages} {3131}
  (\bibinfo {year} {2005})}\BibitemShut {NoStop}%
\bibitem [{\citenamefont {Ishizaki}\ and\ \citenamefont
  {Fleming}(2009)}]{Ishizaki2009}%
  \BibitemOpen
  \bibfield  {author} {\bibinfo {author} {\bibfnamefont {A.}~\bibnamefont
  {Ishizaki}}\ and\ \bibinfo {author} {\bibfnamefont {G.~R.}\ \bibnamefont
  {Fleming}},\ }\href@noop {} {\bibfield  {journal} {\bibinfo  {journal} {Proc.
  Natl. Acad. Sci.}\ }\textbf {\bibinfo {volume} {106}},\ \bibinfo {pages}
  {17255} (\bibinfo {year} {2009})}\BibitemShut {NoStop}%
\bibitem [{\citenamefont {Kreisbeck}\ \emph {et~al.}(2011)\citenamefont
  {Kreisbeck}, \citenamefont {Kramer}, \citenamefont {Rodr\'{\i}guez},\ and\
  \citenamefont {Hein}}]{Kreisbeck2011}%
  \BibitemOpen
  \bibfield  {author} {\bibinfo {author} {\bibfnamefont {C.}~\bibnamefont
  {Kreisbeck}}, \bibinfo {author} {\bibfnamefont {T.}~\bibnamefont {Kramer}},
  \bibinfo {author} {\bibfnamefont {M.}~\bibnamefont {Rodr\'{\i}guez}}, \ and\
  \bibinfo {author} {\bibfnamefont {B.}~\bibnamefont {Hein}},\ }\href@noop {}
  {\bibfield  {journal} {\bibinfo  {journal} {J. Chem. Theory Comput.}\
  }\textbf {\bibinfo {volume} {7}},\ \bibinfo {pages} {2166} (\bibinfo {year}
  {2011})}\BibitemShut {NoStop}%
\bibitem [{\citenamefont {Strunz}\ and\ \citenamefont
  {Gisin}(1999)}]{Strunz1999}%
  \BibitemOpen
  \bibfield  {author} {\bibinfo {author} {\bibfnamefont {W.~T.}\ \bibnamefont
  {Strunz}}\ and\ \bibinfo {author} {\bibfnamefont {N.}~\bibnamefont {Gisin}},\
  }\href@noop {} {\bibfield  {journal} {\bibinfo  {journal} {Phys. Rev. Lett.}\
  }\textbf {\bibinfo {volume} {82}},\ \bibinfo {pages} {1801} (\bibinfo {year}
  {1999})}\BibitemShut {NoStop}%
\bibitem [{\citenamefont {Lacroix}(2008)}]{Lacroix2008}%
  \BibitemOpen
  \bibfield  {author} {\bibinfo {author} {\bibfnamefont {D.}~\bibnamefont
  {Lacroix}},\ }\href@noop {} {\bibfield  {journal} {\bibinfo  {journal} {Phys.
  Rev. E}\ }\textbf {\bibinfo {volume} {77}},\ \bibinfo {pages} {041126}
  (\bibinfo {year} {2008})}\BibitemShut {NoStop}%
\bibitem [{\citenamefont {Li}\ \emph {et~al.}(2011)\citenamefont {Li},
  \citenamefont {Shao},\ and\ \citenamefont {Wang}}]{Li2011}%
  \BibitemOpen
  \bibfield  {author} {\bibinfo {author} {\bibfnamefont {H.}~\bibnamefont
  {Li}}, \bibinfo {author} {\bibfnamefont {J.}~\bibnamefont {Shao}}, \ and\
  \bibinfo {author} {\bibfnamefont {S.}~\bibnamefont {Wang}},\ }\href@noop {}
  {\bibfield  {journal} {\bibinfo  {journal} {Phys. Rev. E}\ }\textbf {\bibinfo
  {volume} {84}},\ \bibinfo {pages} {051112} (\bibinfo {year}
  {2011})}\BibitemShut {NoStop}%
\bibitem [{\citenamefont {Orth}\ \emph {et~al.}(2013)\citenamefont {Orth},
  \citenamefont {Imambekov},\ and\ \citenamefont {{Le Hur}}}]{Orth2013}%
  \BibitemOpen
  \bibfield  {author} {\bibinfo {author} {\bibfnamefont {P.~P.}\ \bibnamefont
  {Orth}}, \bibinfo {author} {\bibfnamefont {A.}~\bibnamefont {Imambekov}}, \
  and\ \bibinfo {author} {\bibfnamefont {K.}~\bibnamefont {{Le Hur}}},\
  }\href@noop {} {\bibfield  {journal} {\bibinfo  {journal} {Phys. Rev. B}\
  }\textbf {\bibinfo {volume} {87}},\ \bibinfo {pages} {014305} (\bibinfo
  {year} {2013})}\BibitemShut {NoStop}%
\bibitem [{\citenamefont {Wang}\ and\ \citenamefont {Thoss}(2008)}]{Wang2008}%
  \BibitemOpen
  \bibfield  {author} {\bibinfo {author} {\bibfnamefont {H.}~\bibnamefont
  {Wang}}\ and\ \bibinfo {author} {\bibfnamefont {M.}~\bibnamefont {Thoss}},\
  }\href@noop {} {\bibfield  {journal} {\bibinfo  {journal} {New J. Phys.}\
  }\textbf {\bibinfo {volume} {10}},\ \bibinfo {pages} {115005} (\bibinfo
  {year} {2008})}\BibitemShut {NoStop}%
\bibitem [{\citenamefont {Schollwöck}(2005)}]{Schollw_ck_2005}%
  \BibitemOpen
  \bibfield  {author} {\bibinfo {author} {\bibfnamefont {U.}~\bibnamefont
  {Schollwöck}},\ }\href@noop {} {\bibfield  {journal} {\bibinfo  {journal}
  {Rev. Mod. Phys.}\ }\textbf {\bibinfo {volume} {77}},\ \bibinfo {pages}
  {259–315} (\bibinfo {year} {2005})}\BibitemShut {NoStop}%
\bibitem [{\citenamefont {Bulla}\ \emph {et~al.}(2008)\citenamefont {Bulla},
  \citenamefont {Costi},\ and\ \citenamefont {Pruschke}}]{Bulla_2008}%
  \BibitemOpen
  \bibfield  {author} {\bibinfo {author} {\bibfnamefont {R.}~\bibnamefont
  {Bulla}}, \bibinfo {author} {\bibfnamefont {T.}~\bibnamefont {Costi}}, \ and\
  \bibinfo {author} {\bibfnamefont {T.}~\bibnamefont {Pruschke}},\ }\href@noop
  {} {\bibfield  {journal} {\bibinfo  {journal} {Rev. Mod. Phys.}\ }\textbf
  {\bibinfo {volume} {80}},\ \bibinfo {pages} {395–450} (\bibinfo {year}
  {2008})}\BibitemShut {NoStop}%
\bibitem [{\citenamefont {Prior}\ \emph {et~al.}(2010)\citenamefont {Prior},
  \citenamefont {Chin}, \citenamefont {Huelga},\ and\ \citenamefont
  {Plenio}}]{Prior_2010}%
  \BibitemOpen
  \bibfield  {author} {\bibinfo {author} {\bibfnamefont {J.}~\bibnamefont
  {Prior}}, \bibinfo {author} {\bibfnamefont {A.~W.}\ \bibnamefont {Chin}},
  \bibinfo {author} {\bibfnamefont {S.~F.}\ \bibnamefont {Huelga}}, \ and\
  \bibinfo {author} {\bibfnamefont {M.~B.}\ \bibnamefont {Plenio}},\
  }\href@noop {} {\bibfield  {journal} {\bibinfo  {journal} {Phys. Rev. Lett.}\
  }\textbf {\bibinfo {volume} {105}},\ \bibinfo {pages} {050404} (\bibinfo
  {year} {2010})}\BibitemShut {NoStop}%
\bibitem [{\citenamefont {Chin}\ \emph {et~al.}(2011)\citenamefont {Chin},
  \citenamefont {Prior}, \citenamefont {Huelga},\ and\ \citenamefont
  {Plenio}}]{Chin_2011}%
  \BibitemOpen
  \bibfield  {author} {\bibinfo {author} {\bibfnamefont {A.~W.}\ \bibnamefont
  {Chin}}, \bibinfo {author} {\bibfnamefont {J.}~\bibnamefont {Prior}},
  \bibinfo {author} {\bibfnamefont {S.~F.}\ \bibnamefont {Huelga}}, \ and\
  \bibinfo {author} {\bibfnamefont {M.~B.}\ \bibnamefont {Plenio}},\
  }\href@noop {} {\bibfield  {journal} {\bibinfo  {journal} {Phys. Rev. Lett.}\
  }\textbf {\bibinfo {volume} {107}},\ \bibinfo {pages} {160601} (\bibinfo
  {year} {2011})}\BibitemShut {NoStop}%
\bibitem [{\citenamefont {Nalbach}\ \emph {et~al.}(2011)\citenamefont
  {Nalbach}, \citenamefont {Braun},\ and\ \citenamefont
  {Thorwart}}]{Nalbach2011}%
  \BibitemOpen
  \bibfield  {author} {\bibinfo {author} {\bibfnamefont {P.}~\bibnamefont
  {Nalbach}}, \bibinfo {author} {\bibfnamefont {D.}~\bibnamefont {Braun}}, \
  and\ \bibinfo {author} {\bibfnamefont {M.}~\bibnamefont {Thorwart}},\
  }\href@noop {} {\bibfield  {journal} {\bibinfo  {journal} {Phys. Rev. E}\
  }\textbf {\bibinfo {volume} {84}},\ \bibinfo {pages} {041926} (\bibinfo
  {year} {2011})}\BibitemShut {NoStop}%
\bibitem [{\citenamefont {Mostame}\ \emph {et~al.}()\citenamefont {Mostame},
  \citenamefont {Huh}, \citenamefont {Kerman}, \citenamefont {Kreisbeck},
  \citenamefont {Fujita}, \citenamefont {Eisfeld},\ and\ \citenamefont
  {Aspuru-Guzik}}]{Mostame:2014}%
  \BibitemOpen
  \bibfield  {author} {\bibinfo {author} {\bibfnamefont {S.}~\bibnamefont
  {Mostame}}, \bibinfo {author} {\bibfnamefont {J.}~\bibnamefont {Huh}},
  \bibinfo {author} {\bibfnamefont {A.~J.}\ \bibnamefont {Kerman}}, \bibinfo
  {author} {\bibfnamefont {C.}~\bibnamefont {Kreisbeck}}, \bibinfo {author}
  {\bibfnamefont {T.}~\bibnamefont {Fujita}}, \bibinfo {author} {\bibfnamefont
  {A.}~\bibnamefont {Eisfeld}}, \ and\ \bibinfo {author} {\bibfnamefont
  {A.}~\bibnamefont {Aspuru-Guzik}},\ }\href@noop {} {\emph {\bibinfo {title}
  {{Spectroscopy of generalized Holstein model using superconducting circuits,
  to be published}}}}\BibitemShut {NoStop}%
\bibitem [{\citenamefont {Chin}\ \emph {et~al.}(2010)\citenamefont {Chin},
  \citenamefont {Rivas}, \citenamefont {Huelga},\ and\ \citenamefont
  {Plenio}}]{Chin_2010}%
  \BibitemOpen
  \bibfield  {author} {\bibinfo {author} {\bibfnamefont {A.~W.}\ \bibnamefont
  {Chin}}, \bibinfo {author} {\bibfnamefont {A.}~\bibnamefont {Rivas}},
  \bibinfo {author} {\bibfnamefont {S.~F.}\ \bibnamefont {Huelga}}, \ and\
  \bibinfo {author} {\bibfnamefont {M.~B.}\ \bibnamefont {Plenio}},\
  }\href@noop {} {\bibfield  {journal} {\bibinfo  {journal} {J. Math. Phys.}\
  }\textbf {\bibinfo {volume} {51}},\ \bibinfo {pages} {092109} (\bibinfo
  {year} {2010})}\BibitemShut {NoStop}%
\bibitem [{\citenamefont {Wilson}(1975)}]{Wilson_1975}%
  \BibitemOpen
  \bibfield  {author} {\bibinfo {author} {\bibfnamefont {K.}~\bibnamefont
  {Wilson}},\ }\href@noop {} {\bibfield  {journal} {\bibinfo  {journal} {Rev.
  Mod. Phys.}\ }\textbf {\bibinfo {volume} {47}},\ \bibinfo {pages} {773–840}
  (\bibinfo {year} {1975})}\BibitemShut {NoStop}%
\bibitem [{\citenamefont {Garg}\ \emph {et~al.}(1985)\citenamefont {Garg},
  \citenamefont {Onuchic},\ and\ \citenamefont {Ambegaokar}}]{Garg_1985}%
  \BibitemOpen
  \bibfield  {author} {\bibinfo {author} {\bibfnamefont {A.}~\bibnamefont
  {Garg}}, \bibinfo {author} {\bibfnamefont {J.~N.}\ \bibnamefont {Onuchic}}, \
  and\ \bibinfo {author} {\bibfnamefont {V.}~\bibnamefont {Ambegaokar}},\
  }\href@noop {} {\bibfield  {journal} {\bibinfo  {journal} {J. Chem. Phys.}\
  }\textbf {\bibinfo {volume} {83}},\ \bibinfo {pages} {4491} (\bibinfo {year}
  {1985})}\BibitemShut {NoStop}%
\bibitem [{\citenamefont {Tretiak}\ \emph {et~al.}(1996)\citenamefont
  {Tretiak}, \citenamefont {Chernyak},\ and\ \citenamefont
  {Mukamel}}]{Tretiak_1996}%
  \BibitemOpen
  \bibfield  {author} {\bibinfo {author} {\bibfnamefont {S.}~\bibnamefont
  {Tretiak}}, \bibinfo {author} {\bibfnamefont {V.}~\bibnamefont {Chernyak}}, \
  and\ \bibinfo {author} {\bibfnamefont {S.}~\bibnamefont {Mukamel}},\
  }\href@noop {} {\bibfield  {journal} {\bibinfo  {journal} {Chem. Phys.
  Lett.}\ }\textbf {\bibinfo {volume} {259}},\ \bibinfo {pages} {55–61}
  (\bibinfo {year} {1996})}\BibitemShut {NoStop}%
\bibitem [{\citenamefont {Mori}(1965)}]{Mori_1965}%
  \BibitemOpen
  \bibfield  {author} {\bibinfo {author} {\bibfnamefont {H.}~\bibnamefont
  {Mori}},\ }\href@noop {} {\bibfield  {journal} {\bibinfo  {journal} {Prog.
  Theor. Phys.}\ }\textbf {\bibinfo {volume} {34}},\ \bibinfo {pages}
  {399–416} (\bibinfo {year} {1965})}\BibitemShut {NoStop}%
\bibitem [{\citenamefont {Dupuis}(1967)}]{Dupuis_1967}%
  \BibitemOpen
  \bibfield  {author} {\bibinfo {author} {\bibfnamefont {M.}~\bibnamefont
  {Dupuis}},\ }\href@noop {} {\bibfield  {journal} {\bibinfo  {journal} {Prog.
  Theor. Phys.}\ }\textbf {\bibinfo {volume} {37}},\ \bibinfo {pages}
  {502–537} (\bibinfo {year} {1967})}\BibitemShut {NoStop}%
\bibitem [{\citenamefont {Grigolini}\ and\ \citenamefont
  {Parravicini}(1982)}]{Grigolini_1982}%
  \BibitemOpen
  \bibfield  {author} {\bibinfo {author} {\bibfnamefont {P.}~\bibnamefont
  {Grigolini}}\ and\ \bibinfo {author} {\bibfnamefont {G.}~\bibnamefont
  {Parravicini}},\ }\href@noop {} {\bibfield  {journal} {\bibinfo  {journal}
  {Phys. Rev. B}\ }\textbf {\bibinfo {volume} {25}},\ \bibinfo {pages}
  {5180–5187} (\bibinfo {year} {1982})}\BibitemShut {NoStop}%
\bibitem [{\citenamefont {Burghardt}\ \emph {et~al.}(2012)\citenamefont
  {Burghardt}, \citenamefont {Martinazzo},\ and\ \citenamefont
  {Hughes}}]{Burghardt2012}%
  \BibitemOpen
  \bibfield  {author} {\bibinfo {author} {\bibfnamefont {I.}~\bibnamefont
  {Burghardt}}, \bibinfo {author} {\bibfnamefont {R.}~\bibnamefont
  {Martinazzo}}, \ and\ \bibinfo {author} {\bibfnamefont {K.~H.}\ \bibnamefont
  {Hughes}},\ }\href@noop {} {\bibfield  {journal} {\bibinfo  {journal} {J.
  Chem. Phys.}\ }\textbf {\bibinfo {volume} {137}},\ \bibinfo {pages} {144107}
  (\bibinfo {year} {2012})}\BibitemShut {NoStop}%
\bibitem [{\citenamefont {Chenel}\ \emph {et~al.}(2014)\citenamefont {Chenel},
  \citenamefont {Mangaud}, \citenamefont {Burghardt}, \citenamefont {Meier},\
  and\ \citenamefont {Desouter-Lecomte}}]{Chenel2014}%
  \BibitemOpen
  \bibfield  {author} {\bibinfo {author} {\bibfnamefont {A.}~\bibnamefont
  {Chenel}}, \bibinfo {author} {\bibfnamefont {E.}~\bibnamefont {Mangaud}},
  \bibinfo {author} {\bibfnamefont {I.}~\bibnamefont {Burghardt}}, \bibinfo
  {author} {\bibfnamefont {C.}~\bibnamefont {Meier}}, \ and\ \bibinfo {author}
  {\bibfnamefont {M.}~\bibnamefont {Desouter-Lecomte}},\ }\href@noop {}
  {\bibfield  {journal} {\bibinfo  {journal} {J. Chem. Phys.}\ }\textbf
  {\bibinfo {volume} {140}},\ \bibinfo {pages} {044104} (\bibinfo {year}
  {2014})}\BibitemShut {NoStop}%
\bibitem [{\citenamefont {Skinner}\ and\ \citenamefont
  {Hu}(2008)}]{Skinner2008}%
  \BibitemOpen
  \bibfield  {author} {\bibinfo {author} {\bibfnamefont {A.~J.}\ \bibnamefont
  {Skinner}}\ and\ \bibinfo {author} {\bibfnamefont {B.-L.}\ \bibnamefont
  {Hu}},\ }\href@noop {} {\bibfield  {journal} {\bibinfo  {journal} {Phys. Rev.
  B}\ }\textbf {\bibinfo {volume} {78}},\ \bibinfo {pages} {014302} (\bibinfo
  {year} {2008})}\BibitemShut {NoStop}%
\bibitem [{\citenamefont {Iachello}\ and\ \citenamefont
  {Arima}(1987)}]{Iachello_1987}%
  \BibitemOpen
  \bibfield  {author} {\bibinfo {author} {\bibfnamefont {F.}~\bibnamefont
  {Iachello}}\ and\ \bibinfo {author} {\bibfnamefont {A.}~\bibnamefont
  {Arima}},\ }\href@noop {} {\emph {\bibinfo {title} {{The Interacting Boson
  Model}}}}\ (\bibinfo  {publisher} {Cambridge University Press},\ \bibinfo
  {year} {1987})\BibitemShut {NoStop}%
\bibitem [{\citenamefont {Caprio}(2005)}]{Caprio_2005}%
  \BibitemOpen
  \bibfield  {author} {\bibinfo {author} {\bibfnamefont {M.~A.}\ \bibnamefont
  {Caprio}},\ }\href@noop {} {\bibfield  {journal} {\bibinfo  {journal} {J.
  Phys. A: Math. Gen.}\ }\textbf {\bibinfo {volume} {38}},\ \bibinfo {pages}
  {6385–6392} (\bibinfo {year} {2005})}\BibitemShut {NoStop}%
\bibitem [{\citenamefont {Feynman}(1982)}]{Feynman1982}%
  \BibitemOpen
  \bibfield  {author} {\bibinfo {author} {\bibfnamefont {R.~P.}\ \bibnamefont
  {Feynman}},\ }\href@noop {} {\bibfield  {journal} {\bibinfo  {journal} {Int.
  J. Theor. Phys.}\ }\textbf {\bibinfo {volume} {21}},\ \bibinfo {pages} {467}
  (\bibinfo {year} {1982})}\BibitemShut {NoStop}%
\bibitem [{\citenamefont {Feynman}(1986)}]{Feynman1986}%
  \BibitemOpen
  \bibfield  {author} {\bibinfo {author} {\bibfnamefont {R.~P.}\ \bibnamefont
  {Feynman}},\ }\href@noop {} {\bibfield  {journal} {\bibinfo  {journal}
  {Found. Phys.}\ }\textbf {\bibinfo {volume} {16}},\ \bibinfo {pages} {507}
  (\bibinfo {year} {1986})}\BibitemShut {NoStop}%
\bibitem [{\citenamefont {Lloyd}(1996)}]{Lloyd1996}%
  \BibitemOpen
  \bibfield  {author} {\bibinfo {author} {\bibfnamefont {S.}~\bibnamefont
  {Lloyd}},\ }\href@noop {} {\bibfield  {journal} {\bibinfo  {journal}
  {Science}\ }\textbf {\bibinfo {volume} {273}},\ \bibinfo {pages} {1073}
  (\bibinfo {year} {1996})}\BibitemShut {NoStop}%
\bibitem [{\citenamefont {Buluta}\ and\ \citenamefont
  {Nori}(2009)}]{Buluta2009}%
  \BibitemOpen
  \bibfield  {author} {\bibinfo {author} {\bibfnamefont {I.}~\bibnamefont
  {Buluta}}\ and\ \bibinfo {author} {\bibfnamefont {F.}~\bibnamefont {Nori}},\
  }\href@noop {} {\bibfield  {journal} {\bibinfo  {journal} {Science}\ }\textbf
  {\bibinfo {volume} {326}},\ \bibinfo {pages} {108} (\bibinfo {year}
  {2009})}\BibitemShut {NoStop}%
\bibitem [{\citenamefont {Houck}\ \emph {et~al.}(2012)\citenamefont {Houck},
  \citenamefont {Tureci},\ and\ \citenamefont {Koch}}]{Houck2012}%
  \BibitemOpen
  \bibfield  {author} {\bibinfo {author} {\bibfnamefont {A.~A.}\ \bibnamefont
  {Houck}}, \bibinfo {author} {\bibfnamefont {H.~E.}\ \bibnamefont {Tureci}}, \
  and\ \bibinfo {author} {\bibfnamefont {J.}~\bibnamefont {Koch}},\ }\href@noop
  {} {\bibfield  {journal} {\bibinfo  {journal} {Nat. Phys.}\ }\textbf
  {\bibinfo {volume} {8}},\ \bibinfo {pages} {292} (\bibinfo {year}
  {2012})}\BibitemShut {NoStop}%
\bibitem [{\citenamefont {Mostame}\ \emph {et~al.}(2012)\citenamefont
  {Mostame}, \citenamefont {Rebentrost}, \citenamefont {Eisfeld}, \citenamefont
  {Kerman}, \citenamefont {Tsomokos},\ and\ \citenamefont
  {Aspuru-Guzik}}]{Mostame2012}%
  \BibitemOpen
  \bibfield  {author} {\bibinfo {author} {\bibfnamefont {S.}~\bibnamefont
  {Mostame}}, \bibinfo {author} {\bibfnamefont {P.}~\bibnamefont {Rebentrost}},
  \bibinfo {author} {\bibfnamefont {A.}~\bibnamefont {Eisfeld}}, \bibinfo
  {author} {\bibfnamefont {A.~J.}\ \bibnamefont {Kerman}}, \bibinfo {author}
  {\bibfnamefont {D.~I.}\ \bibnamefont {Tsomokos}}, \ and\ \bibinfo {author}
  {\bibfnamefont {A.}~\bibnamefont {Aspuru-Guzik}},\ }\href@noop {} {\bibfield
  {journal} {\bibinfo  {journal} {New J. Phys.}\ }\textbf {\bibinfo {volume}
  {14}},\ \bibinfo {pages} {105013} (\bibinfo {year} {2012})}\BibitemShut
  {NoStop}%
\bibitem [{\citenamefont {Ballester}\ \emph {et~al.}(2012)\citenamefont
  {Ballester}, \citenamefont {Romero}, \citenamefont {Garcia-Ripoll},
  \citenamefont {Deppe},\ and\ \citenamefont {Solano}}]{Ballester2012}%
  \BibitemOpen
  \bibfield  {author} {\bibinfo {author} {\bibfnamefont {D.}~\bibnamefont
  {Ballester}}, \bibinfo {author} {\bibfnamefont {G.}~\bibnamefont {Romero}},
  \bibinfo {author} {\bibfnamefont {J.~J.}\ \bibnamefont {Garcia-Ripoll}},
  \bibinfo {author} {\bibfnamefont {F.}~\bibnamefont {Deppe}}, \ and\ \bibinfo
  {author} {\bibfnamefont {E.}~\bibnamefont {Solano}},\ }\href@noop {}
  {\bibfield  {journal} {\bibinfo  {journal} {Phys. Rev. X}\ }\textbf {\bibinfo
  {volume} {2}},\ \bibinfo {pages} {021007} (\bibinfo {year}
  {2012})}\BibitemShut {NoStop}%
\bibitem [{\citenamefont {Mei}\ \emph {et~al.}(2013)\citenamefont {Mei},
  \citenamefont {Stojanovi\ifmmode~\acute{c}\else \'{c}\fi{}}, \citenamefont
  {Siddiqi},\ and\ \citenamefont {Tian}}]{Mei2013}%
  \BibitemOpen
  \bibfield  {author} {\bibinfo {author} {\bibfnamefont {F.}~\bibnamefont
  {Mei}}, \bibinfo {author} {\bibfnamefont {V.~M.}\ \bibnamefont
  {Stojanovi\ifmmode~\acute{c}\else \'{c}\fi{}}}, \bibinfo {author}
  {\bibfnamefont {I.}~\bibnamefont {Siddiqi}}, \ and\ \bibinfo {author}
  {\bibfnamefont {L.}~\bibnamefont {Tian}},\ }\href@noop {} {\bibfield
  {journal} {\bibinfo  {journal} {Phys. Rev. B}\ }\textbf {\bibinfo {volume}
  {88}},\ \bibinfo {pages} {224502} (\bibinfo {year} {2013})}\BibitemShut
  {NoStop}%
\bibitem [{\citenamefont {Heras}\ \emph {et~al.}(2014)\citenamefont {Heras},
  \citenamefont {Mezzacapo}, \citenamefont {Lamata}, \citenamefont {Filipp},
  \citenamefont {Wallraff},\ and\ \citenamefont {Solano}}]{Heras2014}%
  \BibitemOpen
  \bibfield  {author} {\bibinfo {author} {\bibfnamefont {U.~L.}\ \bibnamefont
  {Heras}}, \bibinfo {author} {\bibfnamefont {A.}~\bibnamefont {Mezzacapo}},
  \bibinfo {author} {\bibfnamefont {L.}~\bibnamefont {Lamata}}, \bibinfo
  {author} {\bibfnamefont {S.}~\bibnamefont {Filipp}}, \bibinfo {author}
  {\bibfnamefont {A.}~\bibnamefont {Wallraff}}, \ and\ \bibinfo {author}
  {\bibfnamefont {E.}~\bibnamefont {Solano}},\ }\href@noop {} {\bibfield
  {journal} {\bibinfo  {journal} {Phys. Rev. Lett.}\ }\textbf {\bibinfo
  {volume} {112}},\ \bibinfo {pages} {200501} (\bibinfo {year}
  {2014})}\BibitemShut {NoStop}%
\bibitem [{\citenamefont {Stojanovi\ifmmode~\acute{c}\else \'{c}\fi{}}\ \emph
  {et~al.}(2014)\citenamefont {Stojanovi\ifmmode~\acute{c}\else \'{c}\fi{}},
  \citenamefont {Vanevi\ifmmode~\acute{c}\else \'{c}\fi{}}, \citenamefont
  {Demler},\ and\ \citenamefont {Tian}}]{Vladimir2014}%
  \BibitemOpen
  \bibfield  {author} {\bibinfo {author} {\bibfnamefont {V.~M.}\ \bibnamefont
  {Stojanovi\ifmmode~\acute{c}\else \'{c}\fi{}}}, \bibinfo {author}
  {\bibfnamefont {M.}~\bibnamefont {Vanevi\ifmmode~\acute{c}\else \'{c}\fi{}}},
  \bibinfo {author} {\bibfnamefont {E.}~\bibnamefont {Demler}}, \ and\ \bibinfo
  {author} {\bibfnamefont {L.}~\bibnamefont {Tian}},\ }\href@noop {} {\bibfield
   {journal} {\bibinfo  {journal} {Phys. Rev. B}\ }\textbf {\bibinfo {volume}
  {89}},\ \bibinfo {pages} {144508} (\bibinfo {year} {2014})}\BibitemShut
  {NoStop}%
\bibitem [{\citenamefont {Roushan}\ and\ \citenamefont {{\em et
  al.}}(2014)}]{Pedram2014}%
  \BibitemOpen
  \bibfield  {author} {\bibinfo {author} {\bibfnamefont {P.}~\bibnamefont
  {Roushan}}\ and\ \bibinfo {author} {\bibnamefont {{\em et al.}}},\
  }\href@noop {} {\bibfield  {journal} {\bibinfo  {journal} {arXiv:1407.1585}\
  } (\bibinfo {year} {2014})}\BibitemShut {NoStop}%
\bibitem [{\citenamefont {Cirac}\ and\ \citenamefont
  {Zoller}(2000)}]{Cirac2000}%
  \BibitemOpen
  \bibfield  {author} {\bibinfo {author} {\bibfnamefont {J.~I.}\ \bibnamefont
  {Cirac}}\ and\ \bibinfo {author} {\bibfnamefont {P.}~\bibnamefont {Zoller}},\
  }\href@noop {} {\bibfield  {journal} {\bibinfo  {journal} {Nature}\ }\textbf
  {\bibinfo {volume} {404}},\ \bibinfo {pages} {579} (\bibinfo {year}
  {2000})}\BibitemShut {NoStop}%
\bibitem [{\citenamefont {Sch\"{u}tzhold}\ \emph {et~al.}(2007)\citenamefont
  {Sch\"{u}tzhold}, \citenamefont {Uhlmann}, \citenamefont {Petersen},
  \citenamefont {Schmitz}, \citenamefont {Friedenauer},\ and\ \citenamefont
  {Sch\"{a}tz}}]{Ralf2007}%
  \BibitemOpen
  \bibfield  {author} {\bibinfo {author} {\bibfnamefont {R.}~\bibnamefont
  {Sch\"{u}tzhold}}, \bibinfo {author} {\bibfnamefont {M.}~\bibnamefont
  {Uhlmann}}, \bibinfo {author} {\bibfnamefont {L.}~\bibnamefont {Petersen}},
  \bibinfo {author} {\bibfnamefont {H.}~\bibnamefont {Schmitz}}, \bibinfo
  {author} {\bibfnamefont {A.}~\bibnamefont {Friedenauer}}, \ and\ \bibinfo
  {author} {\bibfnamefont {T.}~\bibnamefont {Sch\"{a}tz}},\ }\href@noop {}
  {\bibfield  {journal} {\bibinfo  {journal} {Phys. Rev. Lett.}\ }\textbf
  {\bibinfo {volume} {99}},\ \bibinfo {pages} {201301} (\bibinfo {year}
  {2007})}\BibitemShut {NoStop}%
\bibitem [{\citenamefont {Gerritsma}\ \emph {et~al.}(2011)\citenamefont
  {Gerritsma}, \citenamefont {Lanyon}, \citenamefont {Kirchmair}, \citenamefont
  {Z\"ahringer}, \citenamefont {Hempel}, \citenamefont {Casanova},
  \citenamefont {Garc\'ia-Ripoll}, \citenamefont {Solano}, \citenamefont
  {Blatt},\ and\ \citenamefont {F}}]{Gerritsma2011}%
  \BibitemOpen
  \bibfield  {author} {\bibinfo {author} {\bibfnamefont {R.}~\bibnamefont
  {Gerritsma}}, \bibinfo {author} {\bibfnamefont {B.~P.}\ \bibnamefont
  {Lanyon}}, \bibinfo {author} {\bibfnamefont {G.}~\bibnamefont {Kirchmair}},
  \bibinfo {author} {\bibfnamefont {F.}~\bibnamefont {Z\"ahringer}}, \bibinfo
  {author} {\bibfnamefont {C.}~\bibnamefont {Hempel}}, \bibinfo {author}
  {\bibfnamefont {J.}~\bibnamefont {Casanova}}, \bibinfo {author}
  {\bibfnamefont {J.~J.}\ \bibnamefont {Garc\'ia-Ripoll}}, \bibinfo {author}
  {\bibfnamefont {E.}~\bibnamefont {Solano}}, \bibinfo {author} {\bibfnamefont
  {R.}~\bibnamefont {Blatt}}, \ and\ \bibinfo {author} {\bibfnamefont {R.~C.}\
  \bibnamefont {F}},\ }\href@noop {} {\bibfield  {journal} {\bibinfo  {journal}
  {Phys. Rev. Lett.}\ }\textbf {\bibinfo {volume} {106}},\ \bibinfo {pages}
  {060503} (\bibinfo {year} {2011})}\BibitemShut {NoStop}%
\bibitem [{\citenamefont {Casanova}\ \emph
  {et~al.}(2011{\natexlab{a}})\citenamefont {Casanova}, \citenamefont {Lamata},
  \citenamefont {Egusquiza}, \citenamefont {Gerritsma}, \citenamefont {F},
  \citenamefont {Garc\'i-Ripoll},\ and\ \citenamefont
  {Solano}}]{Casanova2011a}%
  \BibitemOpen
  \bibfield  {author} {\bibinfo {author} {\bibfnamefont {J.}~\bibnamefont
  {Casanova}}, \bibinfo {author} {\bibfnamefont {L.}~\bibnamefont {Lamata}},
  \bibinfo {author} {\bibfnamefont {I.~L.}\ \bibnamefont {Egusquiza}}, \bibinfo
  {author} {\bibfnamefont {R.}~\bibnamefont {Gerritsma}}, \bibinfo {author}
  {\bibfnamefont {R.~C.}\ \bibnamefont {F}}, \bibinfo {author} {\bibfnamefont
  {J.~J.}\ \bibnamefont {Garc\'i-Ripoll}}, \ and\ \bibinfo {author}
  {\bibfnamefont {E.}~\bibnamefont {Solano}},\ }\href@noop {} {\bibfield
  {journal} {\bibinfo  {journal} {Phys. Rev. Lett.}\ }\textbf {\bibinfo
  {volume} {107}},\ \bibinfo {pages} {260501} (\bibinfo {year}
  {2011}{\natexlab{a}})}\BibitemShut {NoStop}%
\bibitem [{\citenamefont {Casanova}\ \emph
  {et~al.}(2011{\natexlab{b}})\citenamefont {Casanova}, \citenamefont {Sabin},
  \citenamefont {Le\'on}, \citenamefont {Egusquiza}, \citenamefont {Gerritsma},
  \citenamefont {Roos}, \citenamefont {Garcia-Ripoll},\ and\ \citenamefont
  {Solano}}]{Casanova2011b}%
  \BibitemOpen
  \bibfield  {author} {\bibinfo {author} {\bibfnamefont {J.}~\bibnamefont
  {Casanova}}, \bibinfo {author} {\bibfnamefont {C.}~\bibnamefont {Sabin}},
  \bibinfo {author} {\bibfnamefont {J.}~\bibnamefont {Le\'on}}, \bibinfo
  {author} {\bibfnamefont {I.~L.}\ \bibnamefont {Egusquiza}}, \bibinfo {author}
  {\bibfnamefont {R.}~\bibnamefont {Gerritsma}}, \bibinfo {author}
  {\bibfnamefont {C.~F.}\ \bibnamefont {Roos}}, \bibinfo {author}
  {\bibfnamefont {J.~J.}\ \bibnamefont {Garcia-Ripoll}}, \ and\ \bibinfo
  {author} {\bibfnamefont {E.}~\bibnamefont {Solano}},\ }\href@noop {}
  {\bibfield  {journal} {\bibinfo  {journal} {Phys. Rev. X}\ }\textbf {\bibinfo
  {volume} {1}},\ \bibinfo {pages} {021018} (\bibinfo {year}
  {2011}{\natexlab{b}})}\BibitemShut {NoStop}%
\bibitem [{\citenamefont {Casanova}\ \emph {et~al.}(2012)\citenamefont
  {Casanova}, \citenamefont {Mezzacapo}, \citenamefont {Lamata},\ and\
  \citenamefont {Solano}}]{Casanova2012a}%
  \BibitemOpen
  \bibfield  {author} {\bibinfo {author} {\bibfnamefont {J.}~\bibnamefont
  {Casanova}}, \bibinfo {author} {\bibfnamefont {A.}~\bibnamefont {Mezzacapo}},
  \bibinfo {author} {\bibfnamefont {L.}~\bibnamefont {Lamata}}, \ and\ \bibinfo
  {author} {\bibfnamefont {E.}~\bibnamefont {Solano}},\ }\href@noop {}
  {\bibfield  {journal} {\bibinfo  {journal} {Phys. Rev. Lett.}\ }\textbf
  {\bibinfo {volume} {108}},\ \bibinfo {pages} {190502} (\bibinfo {year}
  {2012})}\BibitemShut {NoStop}%
\bibitem [{\citenamefont {Blatt}\ and\ \citenamefont {Roos}(2012)}]{Blatt2012}%
  \BibitemOpen
  \bibfield  {author} {\bibinfo {author} {\bibfnamefont {R.}~\bibnamefont
  {Blatt}}\ and\ \bibinfo {author} {\bibfnamefont {C.~F.}\ \bibnamefont
  {Roos}},\ }\href@noop {} {\bibfield  {journal} {\bibinfo  {journal} {Nat.
  Phys.}\ }\textbf {\bibinfo {volume} {8}},\ \bibinfo {pages} {277} (\bibinfo
  {year} {2012})}\BibitemShut {NoStop}%
\bibitem [{\citenamefont {Stojanovi\ifmmode~\acute{c}\else \'{c}\fi{}}\ \emph
  {et~al.}(2012)\citenamefont {Stojanovi\ifmmode~\acute{c}\else \'{c}\fi{}},
  \citenamefont {Shi}, \citenamefont {Bruder},\ and\ \citenamefont
  {Cirac}}]{Vladimir2012}%
  \BibitemOpen
  \bibfield  {author} {\bibinfo {author} {\bibfnamefont {V.~M.}\ \bibnamefont
  {Stojanovi\ifmmode~\acute{c}\else \'{c}\fi{}}}, \bibinfo {author}
  {\bibfnamefont {T.}~\bibnamefont {Shi}}, \bibinfo {author} {\bibfnamefont
  {C.}~\bibnamefont {Bruder}}, \ and\ \bibinfo {author} {\bibfnamefont {J.~I.}\
  \bibnamefont {Cirac}},\ }\href@noop {} {\bibfield  {journal} {\bibinfo
  {journal} {Phys. Rev. Lett.}\ }\textbf {\bibinfo {volume} {109}},\ \bibinfo
  {pages} {250501} (\bibinfo {year} {2012})}\BibitemShut {NoStop}%
\bibitem [{\citenamefont {Schindler}\ \emph {et~al.}(2013)\citenamefont
  {Schindler}, \citenamefont {M\"uller}, \citenamefont {Nigg}, \citenamefont
  {Barreiro}, \citenamefont {Martinez}, \citenamefont {Hennrich}, \citenamefont
  {Monz}, \citenamefont {Diehl}, \citenamefont {Zoller},\ and\ \citenamefont
  {Blatt}}]{Schindler2013}%
  \BibitemOpen
  \bibfield  {author} {\bibinfo {author} {\bibfnamefont {P.}~\bibnamefont
  {Schindler}}, \bibinfo {author} {\bibfnamefont {M.}~\bibnamefont {M\"uller}},
  \bibinfo {author} {\bibfnamefont {D.}~\bibnamefont {Nigg}}, \bibinfo {author}
  {\bibfnamefont {J.~T.}\ \bibnamefont {Barreiro}}, \bibinfo {author}
  {\bibfnamefont {E.~A.}\ \bibnamefont {Martinez}}, \bibinfo {author}
  {\bibfnamefont {M.}~\bibnamefont {Hennrich}}, \bibinfo {author}
  {\bibfnamefont {T.}~\bibnamefont {Monz}}, \bibinfo {author} {\bibfnamefont
  {S.}~\bibnamefont {Diehl}}, \bibinfo {author} {\bibfnamefont
  {P.}~\bibnamefont {Zoller}}, \ and\ \bibinfo {author} {\bibfnamefont
  {R.}~\bibnamefont {Blatt}},\ }\href@noop {} {\bibfield  {journal} {\bibinfo
  {journal} {Nat. Phys.}\ }\textbf {\bibinfo {volume} {9}},\ \bibinfo {pages}
  {361} (\bibinfo {year} {2013})}\BibitemShut {NoStop}%
\bibitem [{\citenamefont {Lanyon}\ \emph {et~al.}(2010)\citenamefont {Lanyon},
  \citenamefont {Whitfield}, \citenamefont {Gillett}, \citenamefont {Goggin},
  \citenamefont {Almeida}, \citenamefont {Kassal}, \citenamefont {Biamonte},
  \citenamefont {Mohseni}, \citenamefont {Powell}, \citenamefont {Barbieri},
  \citenamefont {Aspuru-Guzik},\ and\ \citenamefont {White}}]{Lanyon2010}%
  \BibitemOpen
  \bibfield  {author} {\bibinfo {author} {\bibfnamefont {B.~P.}\ \bibnamefont
  {Lanyon}}, \bibinfo {author} {\bibfnamefont {J.~D.}\ \bibnamefont
  {Whitfield}}, \bibinfo {author} {\bibfnamefont {G.~G.}\ \bibnamefont
  {Gillett}}, \bibinfo {author} {\bibfnamefont {M.~E.}\ \bibnamefont {Goggin}},
  \bibinfo {author} {\bibfnamefont {M.~P.}\ \bibnamefont {Almeida}}, \bibinfo
  {author} {\bibfnamefont {I.}~\bibnamefont {Kassal}}, \bibinfo {author}
  {\bibfnamefont {J.~D.}\ \bibnamefont {Biamonte}}, \bibinfo {author}
  {\bibfnamefont {M.}~\bibnamefont {Mohseni}}, \bibinfo {author} {\bibfnamefont
  {B.~J.}\ \bibnamefont {Powell}}, \bibinfo {author} {\bibfnamefont
  {M.}~\bibnamefont {Barbieri}}, \bibinfo {author} {\bibfnamefont
  {A.}~\bibnamefont {Aspuru-Guzik}}, \ and\ \bibinfo {author} {\bibfnamefont
  {A.~G.}\ \bibnamefont {White}},\ }\href@noop {} {\bibfield  {journal}
  {\bibinfo  {journal} {Nat. Chem.}\ }\textbf {\bibinfo {volume} {2}},\
  \bibinfo {pages} {106} (\bibinfo {year} {2010})}\BibitemShut {NoStop}%
\bibitem [{\citenamefont {Szpak}\ and\ \citenamefont
  {Sch\"utzhold}(2011)}]{Ralf2011}%
  \BibitemOpen
  \bibfield  {author} {\bibinfo {author} {\bibfnamefont {N.}~\bibnamefont
  {Szpak}}\ and\ \bibinfo {author} {\bibfnamefont {R.}~\bibnamefont
  {Sch\"utzhold}},\ }\href@noop {} {\bibfield  {journal} {\bibinfo  {journal}
  {Phys. Rev. A}\ }\textbf {\bibinfo {volume} {84}},\ \bibinfo {pages} {050101}
  (\bibinfo {year} {2011})}\BibitemShut {NoStop}%
\bibitem [{\citenamefont {Herrera}\ and\ \citenamefont
  {Krems}(2011)}]{Herrera2011}%
  \BibitemOpen
  \bibfield  {author} {\bibinfo {author} {\bibfnamefont {F.}~\bibnamefont
  {Herrera}}\ and\ \bibinfo {author} {\bibfnamefont {R.~V.}\ \bibnamefont
  {Krems}},\ }\href@noop {} {\bibfield  {journal} {\bibinfo  {journal} {Phys.
  Rev. A}\ }\textbf {\bibinfo {volume} {84}},\ \bibinfo {pages} {051401}
  (\bibinfo {year} {2011})}\BibitemShut {NoStop}%
\bibitem [{\citenamefont {Aspuru-Guzik}\ and\ \citenamefont
  {Walther}(2012)}]{Aspuru2012}%
  \BibitemOpen
  \bibfield  {author} {\bibinfo {author} {\bibfnamefont {A.}~\bibnamefont
  {Aspuru-Guzik}}\ and\ \bibinfo {author} {\bibfnamefont {P.}~\bibnamefont
  {Walther}},\ }\href@noop {} {\bibfield  {journal} {\bibinfo  {journal} {Nat.
  Phys.}\ }\textbf {\bibinfo {volume} {8}},\ \bibinfo {pages} {285} (\bibinfo
  {year} {2012})}\BibitemShut {NoStop}%
\bibitem [{\citenamefont {Bloch}\ \emph {et~al.}(2012)\citenamefont {Bloch},
  \citenamefont {Dalibard},\ and\ \citenamefont {Nascimbene}}]{Bloch2012}%
  \BibitemOpen
  \bibfield  {author} {\bibinfo {author} {\bibfnamefont {I.}~\bibnamefont
  {Bloch}}, \bibinfo {author} {\bibfnamefont {J.}~\bibnamefont {Dalibard}}, \
  and\ \bibinfo {author} {\bibfnamefont {S.}~\bibnamefont {Nascimbene}},\
  }\href@noop {} {\bibfield  {journal} {\bibinfo  {journal} {Nat. Phys.}\
  }\textbf {\bibinfo {volume} {8}},\ \bibinfo {pages} {267} (\bibinfo {year}
  {2012})}\BibitemShut {NoStop}%
\bibitem [{\citenamefont {Du}\ \emph {et~al.}(2010)\citenamefont {Du},
  \citenamefont {Xu}, \citenamefont {Peng}, \citenamefont {Wang}, \citenamefont
  {Wu},\ and\ \citenamefont {Lu}}]{Du2010}%
  \BibitemOpen
  \bibfield  {author} {\bibinfo {author} {\bibfnamefont {J.}~\bibnamefont
  {Du}}, \bibinfo {author} {\bibfnamefont {N.}~\bibnamefont {Xu}}, \bibinfo
  {author} {\bibfnamefont {X.}~\bibnamefont {Peng}}, \bibinfo {author}
  {\bibfnamefont {P.}~\bibnamefont {Wang}}, \bibinfo {author} {\bibfnamefont
  {S.}~\bibnamefont {Wu}}, \ and\ \bibinfo {author} {\bibfnamefont
  {D.}~\bibnamefont {Lu}},\ }\href@noop {} {\bibfield  {journal} {\bibinfo
  {journal} {Phys. Rev. Lett.}\ }\textbf {\bibinfo {volume} {104}},\ \bibinfo
  {pages} {030502} (\bibinfo {year} {2010})}\BibitemShut {NoStop}%
\bibitem [{\citenamefont {Lu}\ \emph {et~al.}(2011)\citenamefont {Lu},
  \citenamefont {Xu}, \citenamefont {Xu}, \citenamefont {Chen}, \citenamefont
  {Gong}, \citenamefont {Peng},\ and\ \citenamefont {Du}}]{Lu2011}%
  \BibitemOpen
  \bibfield  {author} {\bibinfo {author} {\bibfnamefont {D.}~\bibnamefont
  {Lu}}, \bibinfo {author} {\bibfnamefont {N.}~\bibnamefont {Xu}}, \bibinfo
  {author} {\bibfnamefont {R.}~\bibnamefont {Xu}}, \bibinfo {author}
  {\bibfnamefont {H.}~\bibnamefont {Chen}}, \bibinfo {author} {\bibfnamefont
  {J.}~\bibnamefont {Gong}}, \bibinfo {author} {\bibfnamefont {X.}~\bibnamefont
  {Peng}}, \ and\ \bibinfo {author} {\bibfnamefont {J.}~\bibnamefont {Du}},\
  }\href@noop {} {\bibfield  {journal} {\bibinfo  {journal} {Phys. Rev. Lett.}\
  }\textbf {\bibinfo {volume} {107}},\ \bibinfo {pages} {020501} (\bibinfo
  {year} {2011})}\BibitemShut {NoStop}%
\bibitem [{\citenamefont {Zhang}\ \emph {et~al.}(2011)\citenamefont {Zhang},
  \citenamefont {Yung}, \citenamefont {Laflamme}, \citenamefont
  {Aspuru-Guzik},\ and\ \citenamefont {Baugh}}]{Zhang2011}%
  \BibitemOpen
  \bibfield  {author} {\bibinfo {author} {\bibfnamefont {J.}~\bibnamefont
  {Zhang}}, \bibinfo {author} {\bibfnamefont {M.-H.}\ \bibnamefont {Yung}},
  \bibinfo {author} {\bibfnamefont {R.}~\bibnamefont {Laflamme}}, \bibinfo
  {author} {\bibfnamefont {A.}~\bibnamefont {Aspuru-Guzik}}, \ and\ \bibinfo
  {author} {\bibfnamefont {J.}~\bibnamefont {Baugh}},\ }\href@noop {}
  {\bibfield  {journal} {\bibinfo  {journal} {Nat. Commun.}\ }\textbf {\bibinfo
  {volume} {3}} (\bibinfo {year} {2011})}\BibitemShut {NoStop}%
\bibitem [{\citenamefont {Feng}\ \emph {et~al.}(2013)\citenamefont {Feng},
  \citenamefont {Lu}, \citenamefont {Hao}, \citenamefont {Zhang},\ and\
  \citenamefont {Long}}]{Feng2013}%
  \BibitemOpen
  \bibfield  {author} {\bibinfo {author} {\bibfnamefont {G.-R.}\ \bibnamefont
  {Feng}}, \bibinfo {author} {\bibfnamefont {Y.}~\bibnamefont {Lu}}, \bibinfo
  {author} {\bibfnamefont {L.}~\bibnamefont {Hao}}, \bibinfo {author}
  {\bibfnamefont {F.-H.}\ \bibnamefont {Zhang}}, \ and\ \bibinfo {author}
  {\bibfnamefont {G.-L.}\ \bibnamefont {Long}},\ }\href@noop {} {\bibfield
  {journal} {\bibinfo  {journal} {Sci. Rep.}\ }\textbf {\bibinfo {volume} {3}}
  (\bibinfo {year} {2013})}\BibitemShut {NoStop}%
\bibitem [{\citenamefont {Sch\"{u}tzhold}\ and\ \citenamefont
  {Mostame}(2005)}]{Mostame2004}%
  \BibitemOpen
  \bibfield  {author} {\bibinfo {author} {\bibfnamefont {R.}~\bibnamefont
  {Sch\"{u}tzhold}}\ and\ \bibinfo {author} {\bibfnamefont {S.}~\bibnamefont
  {Mostame}},\ }\href@noop {} {\bibfield  {journal} {\bibinfo  {journal} {JETP
  Letters}\ }\textbf {\bibinfo {volume} {82}},\ \bibinfo {pages} {248}
  (\bibinfo {year} {2005})}\BibitemShut {NoStop}%
\bibitem [{\citenamefont {Mostame}\ and\ \citenamefont
  {Sch\"{u}tzhold}(2008)}]{Mostame2008}%
  \BibitemOpen
  \bibfield  {author} {\bibinfo {author} {\bibfnamefont {S.}~\bibnamefont
  {Mostame}}\ and\ \bibinfo {author} {\bibfnamefont {R.}~\bibnamefont
  {Sch\"{u}tzhold}},\ }\href@noop {} {\bibfield  {journal} {\bibinfo  {journal}
  {Phys. Rev. Lett.}\ }\textbf {\bibinfo {volume} {101}},\ \bibinfo {pages}
  {220501} (\bibinfo {year} {2008})}\BibitemShut {NoStop}%
\bibitem [{\citenamefont {Chernyak}\ and\ \citenamefont
  {Mukamel}(1996)}]{Chernyak1996}%
  \BibitemOpen
  \bibfield  {author} {\bibinfo {author} {\bibfnamefont {V.}~\bibnamefont
  {Chernyak}}\ and\ \bibinfo {author} {\bibfnamefont {S.}~\bibnamefont
  {Mukamel}},\ }\href@noop {} {\bibfield  {journal} {\bibinfo  {journal} {J.
  Chem. Phys.}\ }\textbf {\bibinfo {volume} {105}},\ \bibinfo {pages} {4565}
  (\bibinfo {year} {1996})}\BibitemShut {NoStop}%
\bibitem [{\citenamefont {Chou}\ \emph {et~al.}(2008)\citenamefont {Chou},
  \citenamefont {Yu},\ and\ \citenamefont {Hu}}]{Chou2008}%
  \BibitemOpen
  \bibfield  {author} {\bibinfo {author} {\bibfnamefont {C.-H.}\ \bibnamefont
  {Chou}}, \bibinfo {author} {\bibfnamefont {T.}~\bibnamefont {Yu}}, \ and\
  \bibinfo {author} {\bibfnamefont {B.~L.}\ \bibnamefont {Hu}},\ }\href@noop {}
  {\bibfield  {journal} {\bibinfo  {journal} {Phys. Rev. E}\ }\textbf {\bibinfo
  {volume} {77}},\ \bibinfo {pages} {011112} (\bibinfo {year}
  {2008})}\BibitemShut {NoStop}%
\bibitem [{\citenamefont {Vojta}\ \emph {et~al.}(2005)\citenamefont {Vojta},
  \citenamefont {Tong},\ and\ \citenamefont {Bullaik}}]{Vojta_2005}%
  \BibitemOpen
  \bibfield  {author} {\bibinfo {author} {\bibfnamefont {M.}~\bibnamefont
  {Vojta}}, \bibinfo {author} {\bibfnamefont {N.-H.}\ \bibnamefont {Tong}}, \
  and\ \bibinfo {author} {\bibfnamefont {R.}~\bibnamefont {Bullaik}},\
  }\href@noop {} {\bibfield  {journal} {\bibinfo  {journal} {Phys. Rev. Lett.}\
  }\textbf {\bibinfo {volume} {94}},\ \bibinfo {pages} {070604} (\bibinfo
  {year} {2005})}\BibitemShut {NoStop}%
\bibitem [{\citenamefont {Fujita}\ \emph {et~al.}(2012)\citenamefont {Fujita},
  \citenamefont {Brookes}, \citenamefont {Saikin},\ and\ \citenamefont
  {Aspuru-Guzik}}]{Fujita2012}%
  \BibitemOpen
  \bibfield  {author} {\bibinfo {author} {\bibfnamefont {T.}~\bibnamefont
  {Fujita}}, \bibinfo {author} {\bibfnamefont {J.~C.}\ \bibnamefont {Brookes}},
  \bibinfo {author} {\bibfnamefont {S.~K.}\ \bibnamefont {Saikin}}, \ and\
  \bibinfo {author} {\bibfnamefont {A.}~\bibnamefont {Aspuru-Guzik}},\
  }\href@noop {} {\bibfield  {journal} {\bibinfo  {journal} {J. Phys. Chem.
  Lett.}\ }\textbf {\bibinfo {volume} {3}},\ \bibinfo {pages} {2357} (\bibinfo
  {year} {2012})}\BibitemShut {NoStop}%
\bibitem [{\citenamefont {Fujita}\ \emph {et~al.}(2014)\citenamefont {Fujita},
  \citenamefont {Huh}, \citenamefont {Saikin}, \citenamefont {Brookes},\ and\
  \citenamefont {Aspuru-Guzik}}]{Fujita2014}%
  \BibitemOpen
  \bibfield  {author} {\bibinfo {author} {\bibfnamefont {T.}~\bibnamefont
  {Fujita}}, \bibinfo {author} {\bibfnamefont {J.}~\bibnamefont {Huh}},
  \bibinfo {author} {\bibfnamefont {S.~K.}\ \bibnamefont {Saikin}}, \bibinfo
  {author} {\bibfnamefont {J.~C.}\ \bibnamefont {Brookes}}, \ and\ \bibinfo
  {author} {\bibfnamefont {A.}~\bibnamefont {Aspuru-Guzik}},\ }\href@noop {}
  {\bibfield  {journal} {\bibinfo  {journal} {Photosynth. Res.}\ }\textbf
  {\bibinfo {volume} {120}},\ \bibinfo {pages} {273} (\bibinfo {year}
  {2014})}\BibitemShut {NoStop}%
\bibitem [{\citenamefont {Guo}(2012)}]{Guo:2012}%
  \BibitemOpen
  \bibfield  {author} {\bibinfo {author} {\bibfnamefont {C.}~\bibnamefont
  {Guo}},\ }\emph {\bibinfo {title} {Using Density Matrix Renormalization Group
  to Study Open Quantum Systems}},\ \href@noop {} {Ph.D. thesis} (\bibinfo
  {year} {2012})\BibitemShut {NoStop}%
\bibitem [{\citenamefont {Valleau}\ \emph {et~al.}(2012)\citenamefont
  {Valleau}, \citenamefont {Eisfeld},\ and\ \citenamefont
  {Aspuru-Guzik}}]{Valleau_2012}%
  \BibitemOpen
  \bibfield  {author} {\bibinfo {author} {\bibfnamefont {S.}~\bibnamefont
  {Valleau}}, \bibinfo {author} {\bibfnamefont {A.}~\bibnamefont {Eisfeld}}, \
  and\ \bibinfo {author} {\bibfnamefont {A.}~\bibnamefont {Aspuru-Guzik}},\
  }\href@noop {} {\bibfield  {journal} {\bibinfo  {journal} {J. Chem. Phys.}\
  }\textbf {\bibinfo {volume} {137}},\ \bibinfo {pages} {224103} (\bibinfo
  {year} {2012})}\BibitemShut {NoStop}%
\bibitem [{\citenamefont {Hughes}\ \emph {et~al.}(2009)\citenamefont {Hughes},
  \citenamefont {Christ},\ and\ \citenamefont {Burghardt}}]{Hughes2009}%
  \BibitemOpen
  \bibfield  {author} {\bibinfo {author} {\bibfnamefont {K.~H.}\ \bibnamefont
  {Hughes}}, \bibinfo {author} {\bibfnamefont {C.~D.}\ \bibnamefont {Christ}},
  \ and\ \bibinfo {author} {\bibfnamefont {I.}~\bibnamefont {Burghardt}},\
  }\href@noop {} {\bibfield  {journal} {\bibinfo  {journal} {J. Chem. Phys.}\
  }\textbf {\bibinfo {volume} {131}},\ \bibinfo {pages} {024109} (\bibinfo
  {year} {2009})}\BibitemShut {NoStop}%
\bibitem [{\citenamefont {Golub}\ and\ \citenamefont
  {Van~Loan}(1996)}]{Golub:1996}%
  \BibitemOpen
  \bibfield  {author} {\bibinfo {author} {\bibfnamefont {G.~H.}\ \bibnamefont
  {Golub}}\ and\ \bibinfo {author} {\bibfnamefont {C.~F.}\ \bibnamefont
  {Van~Loan}},\ }\href@noop {} {\emph {\bibinfo {title} {Matrix Computations
  (3rd Ed.)}}}\ (\bibinfo  {publisher} {Johns Hopkins University Press},\
  \bibinfo {address} {Baltimore, MD, USA},\ \bibinfo {year} {1996})\BibitemShut
  {NoStop}%
\bibitem [{\citenamefont {MATLAB}(2013)}]{MATLAB:2013}%
  \BibitemOpen
  \bibfield  {author} {\bibinfo {author} {\bibnamefont {MATLAB}},\ }\href@noop
  {} {\emph {\bibinfo {title} {version R2013a}}}\ (\bibinfo  {publisher} {The
  MathWorks Inc.},\ \bibinfo {address} {Natick, Massachusetts},\ \bibinfo
  {year} {2013})\BibitemShut {NoStop}%
\bibitem [{\citenamefont {Blankenship}(2002)}]{blankenshipbook}%
  \BibitemOpen
  \bibfield  {author} {\bibinfo {author} {\bibfnamefont {R.~E.}\ \bibnamefont
  {Blankenship}},\ }\href@noop {} {\emph {\bibinfo {title} {{Molecular
  Mechanisms of Photosynthesis}}}}\ (\bibinfo  {publisher} {World Scientific},\
  \bibinfo {address} {London},\ \bibinfo {year} {2002})\BibitemShut {NoStop}%
\bibitem [{\citenamefont {Businger}(1969)}]{Businger1969}%
  \BibitemOpen
  \bibfield  {author} {\bibinfo {author} {\bibfnamefont {P.~A.}\ \bibnamefont
  {Businger}},\ }\href@noop {} {\bibfield  {journal} {\bibinfo  {journal}
  {Math. Comput.}\ }\textbf {\bibinfo {volume} {23}},\ \bibinfo {pages} {819}
  (\bibinfo {year} {1969})}\BibitemShut {NoStop}%
\end{thebibliography}
%

\end{document}